\documentclass[fleqn, usenatbib, twocolumn]{aastex631}

\usepackage{newtxtext,newtxmath, textcomp}
\usepackage{booktabs}
\usepackage{csvsimple}
\usepackage{listings}
\usepackage{multirow}
\usepackage[T1]{fontenc}

\usepackage{enumitem}

\usepackage{graphicx}	
\usepackage{bm}	
\usepackage{array}
\usepackage{hhline}
\usepackage{rotating}
\usepackage{longtable}
\usepackage[flushleft]{threeparttable}
\usepackage{float}
\usepackage{soul}

\usepackage{makecell}
\usepackage{afterpage}
\usepackage{xspace}

\def\arcmin {$^\prime$}
\def\arcsec{$^{\prime\prime}$}

\def\ha {{H$\alpha$}\xspace}

\def\sigha{$\sigma(\mathrm{H}\alpha)$}
\renewcommand{\thefootnote}{\arabic{footnote}}

\begin{document}

\title[The MDW \ha Sky Survey: Data Release 1]{The MDW H$\alpha$ Sky Survey: Data Release 1}

\author[0009-0004-7905-1755]{Noor Aftab}
\affiliation{Department of Astronomy, Columbia University \\ 550 W 120th Street \\ 
New York, NY 10027, USA}

\author[0009-0008-5025-9818]{Xunhe (Andrew) Zhang}
\affiliation{Department of Astronomy, Columbia University \\ 550 W 120th Street \\ 
New York, NY 10027, USA}

\author[0000-0003-2692-2321]{Sean Walker}
\affiliation{David Mittelman Observatory \\ 26 Serenity Street \\ Mayhill, NM 88339, USA}

\author[0000-0003-1235-7173]{Dennis di Cicco}
\affiliation{David Mittelman Observatory \\ 26 Serenity Street \\ Mayhill, NM 88339, USA}

\author[0000-0002-2639-2001]{David R. Mittelman}
\affiliation{David Mittelman Observatory \\ 26 Serenity Street \\ Mayhill, NM 88339, USA}

\author[0000-0001-5194-4384]{Sanya Gupta}
\affiliation{Barnard College, Columbia University \\
3009 Broadway, New York, NY 10027, USA}
\affiliation{Department of Physics, Stanford University, Stanford, CA 94305, USA}

\author[0000-0002-6561-9002]{Andrew~K.~Saydjari}
\altaffiliation{Hubble Fellow}
\affiliation{Department of Astrophysical Sciences, Princeton University,
Princeton, NJ 08544 USA}
\email{aksaydjari@gmail.com}

\author[0000-0002-1129-1873]{Mary Putman}
\affiliation{Department of Astronomy, Columbia University \\ 550 W 120th Street \\ 
New York, NY 10027, USA}

\author[0000-0003-2666-4430]{David Schiminovich}
\affiliation{Department of Astronomy, Columbia University \\ 550 W 120th Street \\ 
New York, NY 10027, USA}

\correspondingauthor{Noor Aftab}
\email{mdw-survey@columbia.edu}

\keywords{Surveys -- Catalogs -- Interstellar medium -- H II regions -- Emission line stars }

\begin{abstract}
 The Mittelman-di Cicco-Walker (MDW) \ha Sky Survey is an autonomously-operated all-sky narrow-band (3nm) \ha imaging survey. The survey was founded by amateur astronomers and the northern sky (Decl. $ \geq 0^\circ$) is presented here in its second stage of refinement for academic use. Each 3.6 deg$^2$ MDW field has 12 20-minute individual exposures with a pixel scale of 3.2\arcsec, a typical PSF of 6\arcsec, and a stack point source depth of 16-17 magnitudes. The northern MDW Survey Data Release 1 (DR1) includes: calibrated and raw mean and individual images, star-removed mean fields, and point source catalogs for all images matched to Data Release 1 of the Panoramic Survey Telescope and Rapid Response System (Pan-STARRS1) and the INT Galactic Plane Survey (IGAPS).\footnote[1]{All DR1 components are available at \url{https://mdw.astro.columbia.edu}. The catalogs are also made available in the AAS Journals Zenodo repository:\dataset[doi:10.5281/zenodo.17307324]{https://doi.org/10.5281/zenodo.17307324}} Our initial study of \ha filament widths finds a typical FWHM of 30-45\arcsec~ in the Lyra region. The matched catalogs (with a median match distance of \textasciitilde0.5\arcsec), combined with our distinctive narrow-band photometry, are used to identify \ha variable and excess sources.  These initial studies highlight some of the many scientific uses of the MDW \ha survey.
\end{abstract}

\keywords{Sky surveys -- Amateur astronomy -- Astronomy data reduction -- H II regions -- Narrow band photometry}

 
\section{Introduction} 
\label{sec:intro}

Hydrogen Balmer alpha (\ha) emission is a popular tracer of ionized hydrogen in the envelopes of stars and stellar systems, within a galaxy's interstellar medium (ISM), and in denser regions of the fainter circumgalactic medium (CGM). When clouds of neutral hydrogen are ionized by energetic processes like young stellar activity, supernovae, and stellar winds, the freed electrons may recombine with ionized hydrogen to emit radiation at the specific wavelength of \ha \citep[e.g.,][]{reynolds84, warm_ionized_medium}.  Since Hydrogen is the most abundant element in the universe and \ha has a relatively high transition probability, this line can be used as an observational tracer to probe a number of physical processes.

\ha has been the focus of numerous past sky surveys, which have been used to study nebular regions and point-sources in the Milky Way, Magellanic Clouds, and the Local Volume. The designs of these surveys have traded off competing scientific goals, focusing on either low-surface brightness sensitivity to \ha over wide angular extent, or deep point-source sensitivity to stellar or compact sources. Surveys with wide, diffuse coverage include narrow-band SHaSSA \citep{SHASSA}, VTSS \citep{VTSS} and the Fabry-Perot based WH$\alpha$M \citep{NORTHERN_WHAM,SOUTHERN_WHAM}; the latter being a spectroscopic survey with degree-scale angular resolution. These three surveys are combined to produce the arcminute-resolution composite map in \citet{finkbeiner2003}, and each has been important for tracing the warm-ionized component of the Milky Way ISM, and its connection to local star-forming regions and other energetic phenomena. Surveys yielding fine \ha narrow-band point source photometry include IGAPS (covering 1860 deg$^2$; 4.5\% of the whole sky; \cite{IGAPS}), Super-COSMOS (4000 deg$^2$; 10\%; \cite{SuperCOSMOS}), VPHAS$+$ (2000 deg$^2$; 4.8\%; \cite{VPHAS}), and more recently, the J-PLUS (8500 deg$^2$; 20\%; \cite{JPLUS}) and S-PLUS (9300 deg$^2$; 23\%; \cite{SPLUS}) multi-narrowband surveys. These campaigns have yielded numerous detections of accreting and symbiotic binaries. New insights have been gleaned from the \ha observations of single stars and their envelopes, in both early and late stages of evolution, and in periods of activity.

A new generation of targeted spectroscopic surveys includes the Sloan Digital Sky Survey V (SDSS-V) Local Volume Mapper (LVM), and detailed surveys carried out with IFUs on other meter-class telescopes (e.g. MUSE, KCWI, CH$\alpha$S). The LVM - examining the Milky Way's interstellar medium and the energy inflows and outflows of the Milky Way subgroup \citep{local_volume_mapper} - has already made intriguing observations of line ratios across the Orion Molecular Complex \citep{lvm_orion}. These images have a resolution of 35.3\arcsec, alluding to interesting prospects when combined with MDW data (\textasciitilde6\arcsec). Of additional note is the Dragonfly Spectral Line Mapper, which will study the CGM of our Local Volume through numerous narrow bandpass filters, including \ha \citep{dragonfly}.

Founded and operated by three amateur astrophotographers, the MDW Survey fills an important, unexplored niche among \ha surveys of high-angular resolution (pixel scale of 3.2\arcsec, median point spread function (PSF) full width at half maximum (FWHM) of \textasciitilde 6\arcsec) and (eventual) all-sky coverage. The MDW Survey thus provides a unique avenue to study \ha-emitting point sources and extended structures, and has already proven useful in a number of projects examining filaments, supernovae remnants, and emission-line objects: \cite{ursa_major_arc, galactic_halo_snr_sharp_ha_filaments, supernova_remnant_cepheus, uv_optical_emission_high_latitude_snr, m31_oiii_emission, colin_smith_cloud}. 

Filamentary structure is found throughout the ISM and is often associated with star formation and feedback events \citep[e.g.,][]{joubaud19,dewangan23}. The properties of atomic, molecular, and dusty filaments in the Milky Way's ISM have been well studied and measured by several groups \citep[e.g.,][]{heyer16,kalberla16,malinen16}. In particular, the \texttt{FilFinder} tool \citep{Koch:2015dc} has been adapted to detect filaments in three dimensions and has been used to identify diffuse atomic hydrogen filaments (HI) in the ISM  \citep[Putman et al. 2025]{kim23}. In regions with high Galactic latitude, HI filaments are often found to be aligned with the Milky Way magnetic field \citep{clark14,kim23}, indicating they are directly associated with a diffuse ionized component.  Closer to the disk of the Galaxy, the diffuse gas is shaped by feedback events and the filaments may trace the edges of shells and other colliding gaseous flows \citep{soler22,colin_smith_cloud}. The filaments are seen in \ha images \citep[e.g.,][]{wareing06, pon14}, but their overall properties have not typically been quantified.


For stellar-focused astronomers, \ha excess and variability in point sources can indicate a broad range of stellar activity types. Young T-Tauri and Herbig Ae/Be stars are known for variable \ha emission, thought to occur because of variable mass accretion from their circumstellar disk \citep{mass_accretion_young_stars}. Chromospheric activity in M stars can trigger powerful \ha flares, and otherwise can present itself as consistent \ha emission \citep{short_term_variability_m_dwarf, sdss_chromospheric_variability}. Such activity can also be used as an analog for magnetic field variability. Additionally, sources with relative \ha excess (not necessarily \ha variable) can be found by comparing their locations in the color-color plane to that of physically similar sources in the color-absolute magnitude diagram \citep{Fratta2021Selection, Fratta2023Spec}. This can signal potentially unexpected physical activity for a source's type, making them interesting targets for follow-up spectroscopy.

\begin{figure}
    \centering
    \includegraphics[width=1\linewidth]{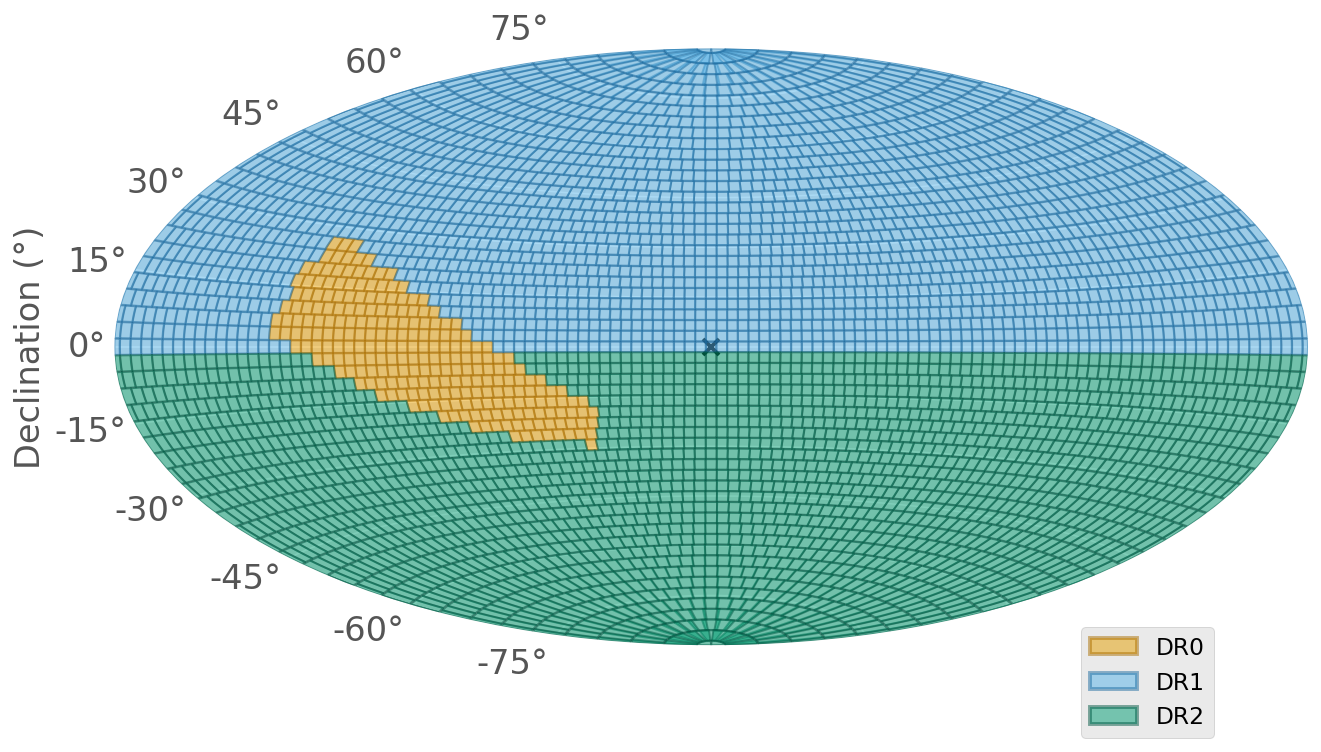}
    \caption{Footprints of the various data releases of the MDW Survey, covering the Orion region (DR0; \cite{MDW_DR0_paper}), the Northern Sky (DR1; detailed here), and eventually the full sky (planned DR2). The cross indicates field 1, with center RA and Dec of $0^\circ$ (RA increases positively to the right).}
    \label{fig:data_releases}
\end{figure}

The first Data Release, Data Release 0 (DR0), of the MDW \ha Sky Survey covers \textasciitilde3100 ${{\rm deg}^2}$ in the Orion region (Dec ranging from $-30^\circ$ to $30^\circ$; yellow region in Figure \ref{fig:data_releases}). We began this release at 0 to indicate its preliminary and exploratory nature. Much of the data collection, reduction, analyses, and products of DR0 are detailed in \cite{MDW_DR0_paper}, along with a comparison of the MDW specifications to existing \ha surveys (Table 1). 

Here, we present the MDW Survey's second Data Release, Data Release 1 (DR1), covering the entire Northern Sky with a footprint spanning $\text{Dec} \geq 0^\circ$, or \textasciitilde20,000 ${{\rm deg}^2}$. This is the blue region in Figure \ref{fig:data_releases}, along with the Northern half of DR0. Section \ref{sec:data_reduction} details the properties of the individual exposures, calibration frames, and a comparison of our \ha filter to IGAPS. Section \ref{sec:pipeline} outlines our processing pipeline for creating the various DR1 components, and Section \ref{sec:reshoots} covers the quality analysis which relies on different aspects of the DR1 pipeline. Section \ref{sec:data_products} provides an overview of the released products, and Section \ref{sec:initial_science} reviews initial science stemming from the DR1 products.

For brevity and clarity, the use of the symbol $\sigma$ in this manuscript always signifies the standard deviation of a property. Several of our plots also include 1$\sigma$, 2$\sigma$ and 3$\sigma$ contour lines, outlining the density of points over three-dimensional visualizations - since we consider our data over different levels of granularity, we indicate in the title of the relevant plots whether these contours describe the density of fields, individual exposures, or point sources.

\section{Observations}
\label{sec:data_reduction}
\begin{figure}
    \centering
    \includegraphics[width=1\linewidth]{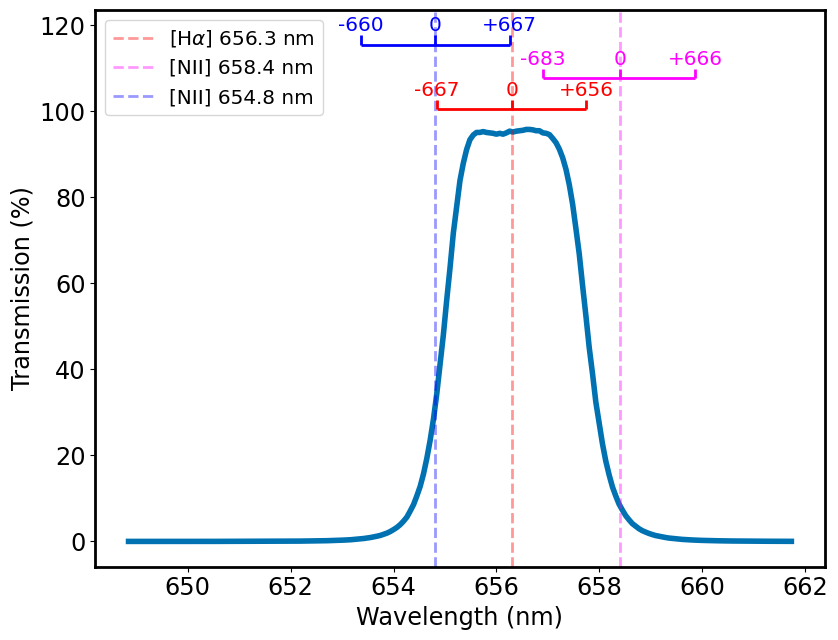}
    \caption{The throughput of the \ha filter used in the MDW Survey, provided by the manufacturer. The band is centered around 656.3nm, with a width of 3nm. We include the velocity ranges for the filter, as well as for the NII lines that may contaminate the \ha band.}
    \label{fig:astrodon_halpha}
\end{figure}

Much of the instrumentation and observation strategy of the MDW Survey is detailed in \cite{MDW_DR0_paper}, but a summary is provided in Table~\ref{tab:key_properties}. A key feature of the MDW Survey is its use of narrow-band \ha filters, which have a bandwidth of 3nm and are commercially manufactured by Astrodon. A standard transmission curve for this filter is provided by Astrodon and is shown in Figure \ref{fig:astrodon_halpha}, along with the filter's associated velocity width. While not investigated for DR1, we expect this curve to be similar (if not the same) for each filter used in the survey's three telescopes. The transmission curve also provides information about potential N II contamination, since the two lines (658.4nm; 654.8nm) are in close proximity to the central \ha wavelength (656.3nm). We include the velocity offsets for these lines in Figure \ref{fig:astrodon_halpha} so users can evaluate possible N II contamination for their objects of interest. 

\begin{table}
    \centering
    \begin{tabular}{>{\raggedright\arraybackslash}p{2cm} >{\raggedright\arraybackslash}p{4cm} }
    \hline
        \multicolumn{1}{l}{\textbf{Property}} & 
        \multicolumn{1}{p{4cm}}{\textbf{Value}}  \\
        \hline
        \multicolumn{1}{l}{Telescope} & 
        \multicolumn{1}{p{4cm}}{Astro-Physics 130GTX}  \\
        
        \multicolumn{1}{l}{Camera} & 
        \multicolumn{1}{p{4cm}}{FLI ProLine 16803}  \\

        \multicolumn{1}{l}{Observing locations} & 
        \multicolumn{1}{p{4cm}}{David Mittelman Observatory (New Mexico, USA); ObsTech SpA (Río Hurtado, Chile)}  \\
        
         
        \multicolumn{1}{l}{Exposure time} & 
        \multicolumn{1}{p{4cm}}{2-4 hours}  \\
        
        \multicolumn{1}{l}{Pixel scale} & 
        \multicolumn{1}{p{4cm}}{3.2 arcsec/pixel}  \\


        
        \multicolumn{1}{l}{Magnitude System} & 
        \multicolumn{1}{p{4cm}}{AB}  \\
        \multicolumn{1}{l}{Individual source depth} & 
        \multicolumn{1}{p{4cm}}{14th-15th mag}  \\
        \multicolumn{1}{l}{Stack source depth} & 
        \multicolumn{1}{p{4cm}}{16th-17th mag}  \\
        
        
        \multicolumn{1}{l}{Field size} & 
        \multicolumn{1}{p{4cm}}{$3.6 \times 3.6$ degrees}  \\
        
        \multicolumn{1}{l}{Survey footprint} & 
        \multicolumn{1}{p{4cm}}{All-sky}  \\
        
        \multicolumn{1}{l}{DR1 footprint} & 
        \multicolumn{1}{p{4cm}}{$\text{Dec} \geq 0^\circ$ (\textasciitilde20,000 ${{\rm deg}^2}$)}  \\

        \multicolumn{1}{p{3.5cm}}{Fraction of fields observed with X unique telescopes} & 
        \multicolumn{1}{p{4cm}}{1: $5.40\%$, 2: $90.7\%$, 3: $3.90\%$}  \\
        
        \hline
    \end{tabular}

    \caption{Properties of Data Release 1 of the MDW \ha Sky Survey.}
    \label{tab:key_properties}
    
\end{table}

\subsection{Individual Frames}
\label{sec:cadence}
Each field in the MDW Survey is imaged roughly 12 times, and each of the 12 images are exposed for 20 minutes. In DR1, we release the 6-12 individual images that are of sufficient quality to generate the mean-combined stack for each field (Section \ref{sec:stacking}), with the eventual goal of releasing 9-12 individual images for all MDW fields. MDW Survey fields are also designed to overlap with each other by 11', or \textasciitilde200 pix, ensuring complete coverage of the release area and allowing us leeway when trimming during the processing pipeline.



\subsubsection{Cadence of observations}

\begin{figure}
    \centering
    \includegraphics[width=1\linewidth]{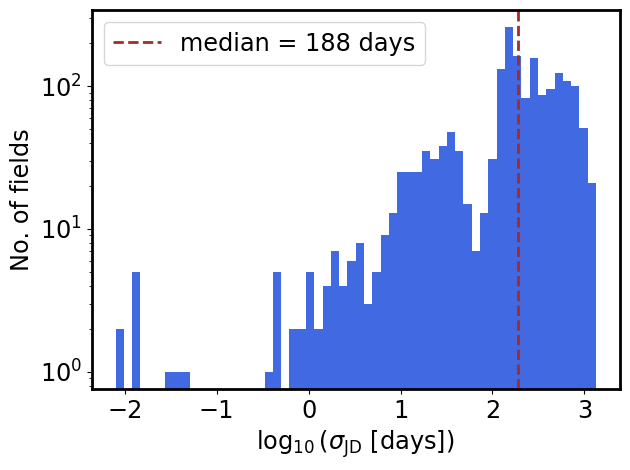}
    \caption{The distribution of variation in exposure observation time for the images that make up each stacked DR1 field.}
    \label{fig:JD_std}
\end{figure}

The DR1 dataset includes the individual, multi-epoch exposures that make up each field's mean-combined stack. It is important to note the irregular cadence of these exposures, particularly for users interested in point-source variability where such a cadence can be a hindrance. The distribution of the standard deviation in exposure observation time ($\sigma_{\mathrm{JD}}$) for DR1 fields is shown in Figure \ref{fig:JD_std}, revealing a median $\sigma_{\mathrm{JD}} = 188$ \ days (dashed red line). 


\begin{figure}
    \centering
    \includegraphics[width=1\linewidth]{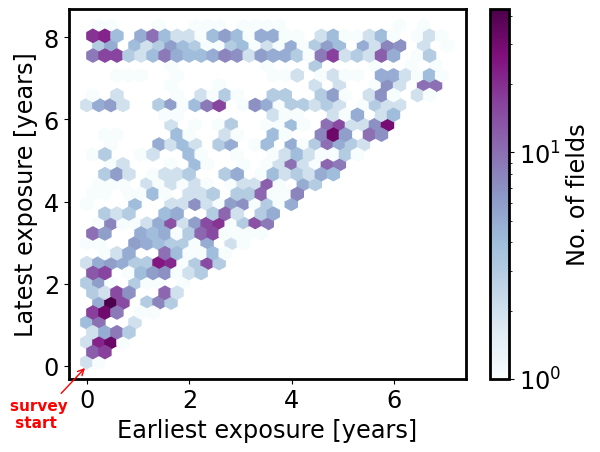}
    \caption{For each DR1 field, a 2D histogram of the observation date of its earliest exposure compared to its latest exposure.}
    \label{fig:JD}
\end{figure}

The range of observation times for the individual images of each field is illustrated in Figure \ref{fig:JD}. Exposures for most fields are taken within a short time frame of each other, as shown by the higher density of points falling in a line of gradient 1, with the median standard deviation among all fields being 9 days. However, there are a number of fields with exposures taken months or years apart, causing the "triangle" spread of Figure \ref{fig:JD}. These gaps in observation may occur because of a variety of reasons: a field's proximity to the moon, a field's accessibility changing over the seasons, scheduling irregularities, or telescopes taken offline for maintenance. In the top left of Figure \ref{fig:JD}, we also see a higher density of fields where the earliest exposure of a field is taken at the beginning of the survey, while the latest exposure is observed more recently. This is the result of the quality control measures discussed in Section \ref{sec:reshoots}. For most of the MDW Survey's duration, fields are scheduled for observation in batches of 2 to 3 consecutive exposures, which allows for momentum in covering the sky. As observations for the MDW Survey come to an end, the priority has shifted to fully completing fields, leading to a greater number of one-off exposures in DR1.

\subsubsection{Stability of exposures}
\label{sec:stability}
\begin{figure*}
    \centering
    \includegraphics[width=0.45\linewidth]{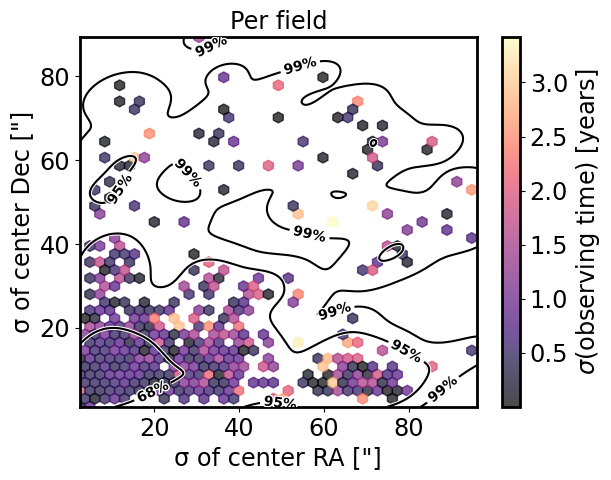}
\hspace{0.05\linewidth}\includegraphics[width=0.45\linewidth]{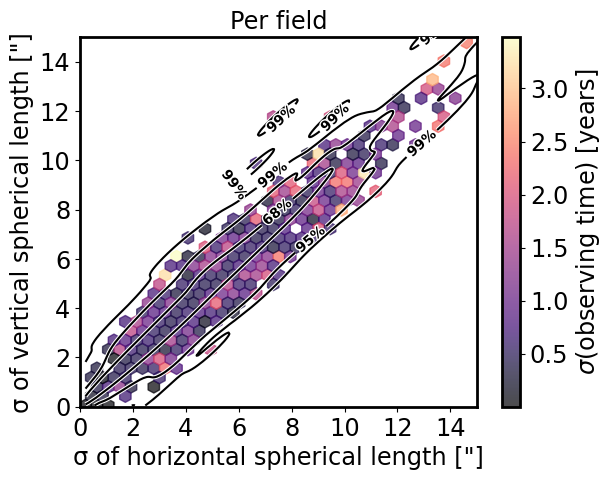}
    \caption{For each DR1 field, we plot the deviation in the exposures center coordinate (left) and the deviation in their horizontal and vertical spherical lengths for the same set of consecutive pixels (right). For both plots, we color the bins by the standard deviation in exposure observing times (also see Figure \ref{fig:JD}). Many outlier fields have either a broad or narrow range in exposure observation dates.}
    \label{fig:stability}
\end{figure*}

In Figure \ref{fig:stability} (left), we compare the center coordinates of the individual exposures for each field. We see that the coordinates are accurate within 25\arcsec, or \textasciitilde8 pixels, in both the RA and Dec directions for the majority of fields, indicating generally stable tracking by our telescope mounts. We also see that a high deviation in center RA does not also suggest high deviation in center Dec. We believe this means that images with higher deviation in their center RA, compared to other images for the same field, may have encountered issues during meridian flips, slewing, or tracking. Issues like these would feasibly affect groups of images taken around the same time, explaining the cluster we see spanning the central RA $\sigma$ of 50\arcsec-75\arcsec. This idea is supported by the histogram's binning, colored by the standard deviation in observing time for the field exposures. In the above-mentioned cluster and the numerous "lone" bins, which are scattered throughout the figure beyond  $\sigma_{\mathrm{center \ RA}} =$ 50\arcsec and $\sigma_{\mathrm{center \ Dec}} =$ 40\arcsec, they tend to show either little deviation in observing time ($\sigma_{\mathrm{JD}} \leq 0.5 \ \mathrm{years}$) or large deviations ($\sigma_{\mathrm{JD}} \geq 2.5 \ \mathrm{years}$).  However, even these outlying fields have center deviations well within the 640\arcsec (200 pixels) overlap margin between fields, so the uniformity of our sky coverage is unchanged. In general, it is good for users to be aware that certain characteristics of the released exposures may differ according to the time elapsed between observations.

In addition to the checks on exposure centers, we also quantify the variation in lens distortion and telescope flexure between individual exposures. To do this, we choose fixed lengths (75\% of the image length) across the horizontal and vertical axes. We convert these lengths to spherical lengths using the astrometric solutions presented in Section \ref{sec:astrometry} to see how much they change between exposures. These deviations are shown in Figure \ref{fig:stability} (right). The exposures for over two-thirds of DR1 fields have spherical lengths varying by less than 10\arcsec (\textasciitilde3 pixels). Most fields fall along a line of gradient 1, indicating that any changes to the spherical length are symmetrical over the image. The consistency we see here in Figure \ref{fig:stability} also illustrates the effect of standardizing each telescope in the MDW Survey to minimize flexure and possible distortion (Appendix A of \cite{MDW_DR0_paper}) .

\subsection{Calibration Frames}
Calibration frames for all three telescopes are acquired automatically using the ACP Expert scheduling software. Flat frames are taken almost on a daily basis, while dark and bias frames are collected every few weeks.  We added darks, dark frames matched in exposure time to the flat fields, to the observing schedule in 2023 when the Southern Hemisphere observations began. Although these constitute only a small portion of the DR1 data, which was primarily acquired before this change, they have been used in the calibration of DR1 reshoots obtained after 2023.

All calibration frames are visually inspected and quantitatively assessed (Section \ref{sec:reshoots}).  They are then combined and applied to the corresponding science frames during pipeline processing (see Section \ref{sec:pipeline}).

\subsection{Comparison of synthetic main-sequence photometry to IGAPS}
\label{sec:synphot}

For consistency with our initial DR0 dataset, we compare to IGAPS when calibrating DR1 photometry (Section \ref{sec:psm_flux_calibration}). However, the IGAPS survey uses a broader filter width (9.5nm), which can result in different photometry for identical sources. To constrain the observational difference between these different filter widths, and to inform how we use the IGAPS photometry in calibrating our own point sources, we calculate synthetic \ha photometry for various dwarf spectral types. 

\begin{figure}
    \centering
    \includegraphics[width=1\linewidth]{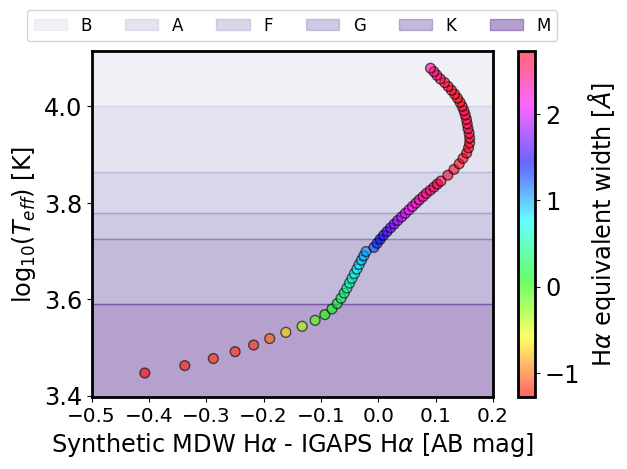}
    \caption{Comparing differences in synthetic photometry of main sequence stars between the MDW Survey and IGAPS \ha filters. The MDW filter width includes less of the molecular absorption lines on the cooler end, and is more dominated by the Balmer absorption line on the hotter end.}
    \label{fig:synphot}
\end{figure}

We simulate observations of unreddened, main sequence stars using the \texttt{synphot} Python API \citep{synphot}. We base "observations" off the high-resolution PHOENIX spectral library \citep{phoenix}, using only stars with solar-metallicity, no $\alpha$-element enhancement, and with logarithmic surface gravity of $3.5 \mathrm{\ cm \ s}^{-2}$. The Isaac Newton Group (ING) of telescopes provides transmission curves for each filter used in ING surveys, and we base the simulations on the transmission curve for the provided IGAPS \ha filter, centered at 6568 \AA \ with a width of 95 \AA. The MDW Survey's narrower filter curve is provided by the Astrodon manufacturer, shown in Figure \ref{fig:astrodon_halpha}.

We compute the difference in H$\alpha$ magnitude between simulated observations in the MDW and IGAPS filters, and plot them against the temperatures provided in the PHOENIX spectra files in Figure \ref{fig:synphot}. The difference in magnitude becomes dramatic for cool M stars, with MDW \ha magnitudes brighter by almost half a magnitude.\footnote{We expect the MDW Survey would not be able to detect the colder end of M stars.} The \ha spectra of M stars are usually contaminated with TiO molecular lines – which makes narrow-band \ha photometry especially dependent on the filter width \citep{VPHAS}. MDW's narrow filter also increases sensitivity to potential \ha emission from an M star. 

Inversely, on the high $T_{\rm eff}$ end of Figure \ref{fig:synphot} (B, A, and F regimes), we see fainter MDW \ha magnitudes than in IGAPS. A-type stars are known to have the strongest Balmer absorption lines, and we see that this regime is where MDW magnitudes become faintest compared to IGAPS, by about 0.15 magnitudes, since our observations become dominated by the strong absorption line. As temperatures veer into the hotter B and cooler F spectral regimes and the absorption line weakens, the difference lowers to \textasciitilde 0.08 magnitudes.

\section{Pipeline}
\label{sec:pipeline}

We develop a pipeline to process the raw individual exposures, which are described in Section \ref{sec:cadence}, into the final data products itemized in Section \ref{sec:data_products}. This pipeline is entirely dependent on open-source software, for better transparency compared to the DR0 dataset, which used proprietary commercial software in a significant amount of the processing. An overview of each step in the pipeline is given below with their corresponding section.

\begin{enumerate}
    \item All calibrated data frames are reprocessed and reapplied to the individual science frames (Section \ref{sec:image_calibration}).
    \item Our astrometry process for each individual frame is optimized to improve distortion and decrease blur in the final stacked images (Section \ref{sec:astrometry} and Section \ref{sec:stacking}).
    \item The individual science frames are analyzed for cosmic-rays and satellite trail contamination (Section \ref{sec:cosmic_ray}).
    \item The cleaned frames are mean-combined for improved signal-to-noise (Section \ref{sec:stacking}).
    \item We extract detections in the individual frames and mean-combined stacks, and create point-source catalogs matched to the Pan-STARRS1 and IGAPS surveys (Section \ref{sec:extraction_matching}).
    \item We calibrate the point-source fluxes relative to the Pan-STARRS1 and IGAPS surveys and incorporate these into the calibrated images (Section \ref{sec:psm_flux_calibration}).
    \item Using our detection catalog as a basis, we infill sources according to their brightness to create star-removed images (Section \ref{sec:star_removal}).
    
\end{enumerate}

\subsection{Individual image calibration}
\label{sec:image_calibration}

In the initial DR0 release, calibration frames were processed and combined using CCDStack and Maxim DL. For DR1, this workflow is transitioned to a Python-based pipeline, utilizing \texttt{astropy} and the affiliated \texttt{ccdproc} package for improved automation and consistency.
Master dark and bias frames are generated by mean-combining the sigma-clipped individual calibration frames. The flat frame processing involves two distinct scenarios based on when the data were acquired.
For data collected prior to the introduction of additional dark frames with exposure times matched to flat fields, or flat darks, the flat frame exposure times differed significantly from the exposure times of the available dark frames. As a result, dark frames were normalized by multiplying the ratio of exposure time of the corresponding flat and the dark frame itself, after subtracting the bias from the flats.
Once flat darks were regularly acquired with the usual dark and bias frames, we could directly perform dark subtraction from flat frames, without the need for separate bias subtraction. After subtracting the dark current, all flat frames taken during the same night are normalized. They are then mean-combined with sigma clipping to create the master flat.

Since the dark frame exposure time matches that of the individual science frames, bias subtraction is not required during dark correction. The master dark frame taken immediately before the science frame is subtracted directly, followed by flat-field correction using the appropriate master flat, yielding an instrumentally calibrated individual science frame.

During calibration, we also extract and standardize FITS header information to ensure uniformity in keyword naming, numerical formatting, units, and other metadata conventions across the dataset. This resolves inconsistencies caused by various factors such as changes in software, hardware, and operational procedures over the course of the survey.

\begin{figure}
    \centering
    \includegraphics[width=1\linewidth]{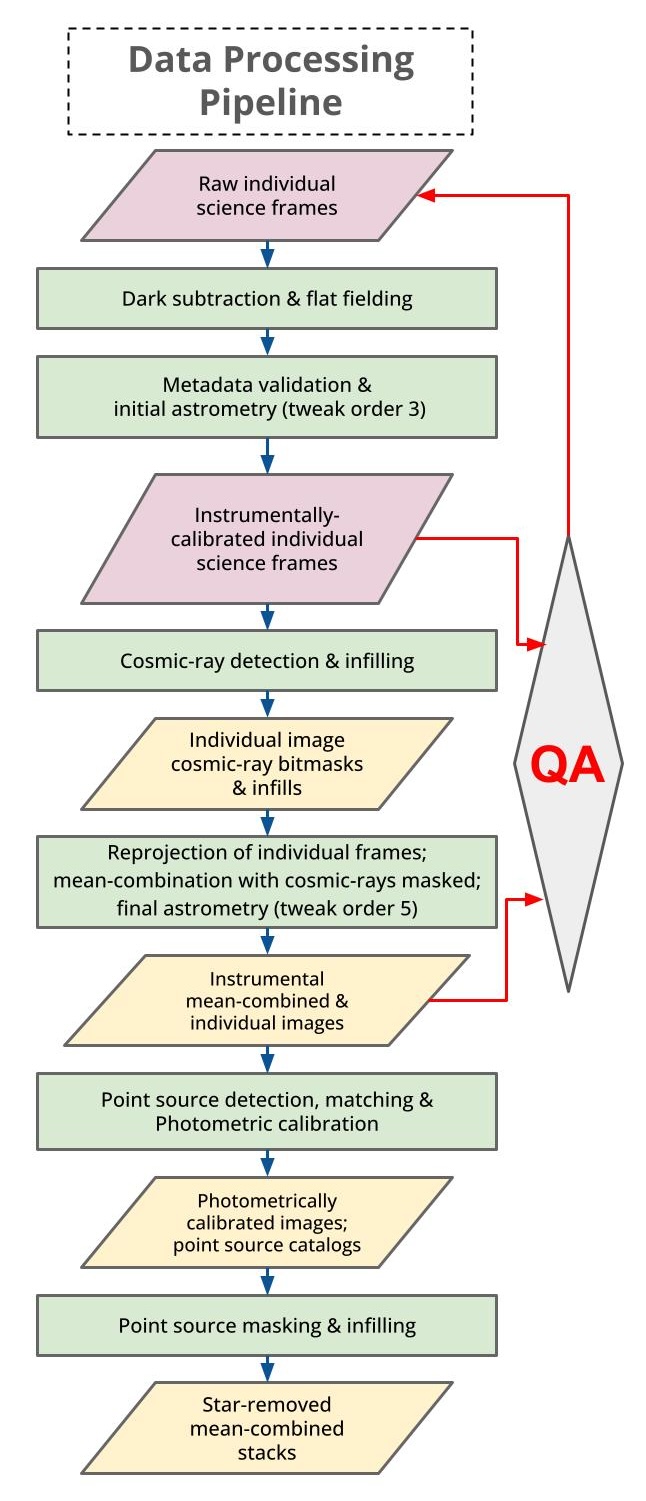}
    \caption{Overview of the data acquisition and processing pipeline. For each step, details are described and discussed in sections \ref{sec:pipeline}, \ref{sec:reshoots}, and  \ref{sec:data_products}. The green boxes indicate processes, while the parallelograms indicate data products (purple for intermediate and unreleased products; yellow for released products) A list of DR1 data products can be found in Table \ref{tab:data_products}. }
    \label{fig:pipeline}
\end{figure}

\subsection{Astrometry}
\label{sec:astrometry}

\begin{figure}
    \centering
    \includegraphics[width=1\linewidth]{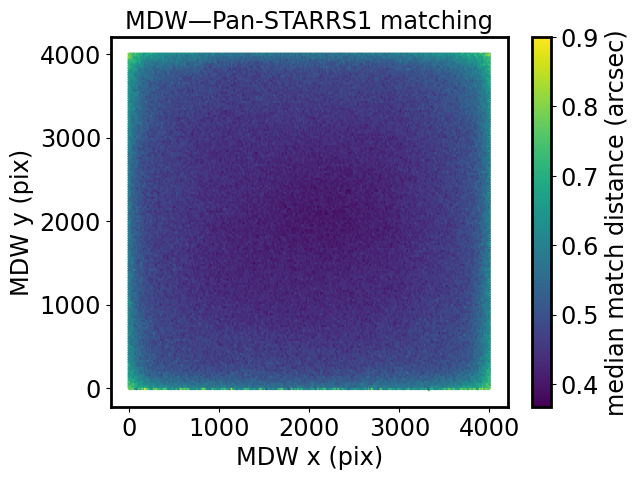}
    \caption{The spatial distribution of match distances between extracted DR1 sources and the Pan-STARRS1 DR1 catalog, binned by the median match distance. We eliminate the "bullseye" pattern seen in Figure 6 of \cite{MDW_DR0_paper}, and increase our astrometric accuracy, illustrated by the smaller range of values in the color bar (up to 2.5" in DR0 vs. 0.9" in DR1).}
    \label{fig:spatial_matches}
\end{figure}

Similar to DR0, we use the \texttt{astrometry.net} software \citep{astrometry_net,astrometry_net_software} to produce the final astrometric solution contained in all image headers included in DR1. Each field image in the MDW Survey has dimensions of $3.6 \times 3.6 $ deg. To ensure the generated solutions fully cover the area of each field, we find it helpful to limit the software to use index files with "skymark diameters" 10\%-20\% the size of an MDW field, by means of a custom configuration file. Since each field has length 213\arcmin, our ideal skymark size ranges from 21\arcmin to 43\arcmin. Our configuration, therefore, forces the software to choose between the 4107 (22\arcmin-30
arcmin), 4108 (30\arcmin-42\arcmin), and 4109 (42\arcmin-60\arcmin) index files provided by the \texttt{astrometry.net} creators. These index files are based off the Tycho2 catalog.


Furthermore, we correct the "bullseye" pattern that was seen in the spatial distribution of coordinate match distances between the Pan-STARRS1 DR1 catalog and the DR0 catalog (Figure 6 in \cite{MDW_DR0_paper}). This pattern was caused by using a lower-order polynomial in the astrometric solutions, which did not fully model the optical distortion of our telescopes, and is eliminated by increasing the polynomial tweak order from 3 to 5 for the final DR1 images.  This is further discussed in Section \ref{sec:stacking}. Figure \ref{fig:spatial_matches} visualizes the updated spatial distribution across fields of match distances between the MDW Survey DR1 stack catalog and the Pan-STARRS1 DR1 catalog. These matches are further discussed in Section \ref{sec:extraction_matching}. We find there is still room for improvement in the DR1 astrometric solutions. This, along with a discussion of Pan-STARRS1 mismatches, is expanded on in Section \ref{sec:dp_catalogs}.


\subsection{Cosmic Ray \& Satellite Trail Identification}
\label{sec:cosmic_ray}

It is important to avoid propagating cosmic rays and satellite trails from the individual images to the final, mean-combined stack, since it may skew our cataloging and point-source flux calibration. We first create a detection algorithm for such artifacts that exploits the fact that DR1 fields are imaged multiple times. If a set of pixels is bright in only one exposure, we assume that it is likely to be a cosmic ray, satellite trail, or other unwanted artifact. A pixel is considered "bright" if its value is greater than 1.8 times the background RMS, which is a hand-tuned factor that appropriately detects most cosmic rays and trails in the MDW Survey's individual images. While this method may mistakenly flag single-frame \ha flares from sources the MDW Survey does not otherwise detect, we expect this to be a small number of events. The single-detection mask is released separately, so users can choose whether they want to apply it when utilizing our individual exposures.

We use the image segmentation API from \texttt{photutils}, to pick out pixels that make up both extended and point-like sources. For each exposure, we iteratively estimate the background with increasing resolution, using the \texttt{photutils} 2D background estimation functions. We find this iterative method helpful to avoid mistakenly detecting "sources" in areas of high nebulosity. In each iteration, we look for sources with decreasing widths and pixel values of at least $1.8 \times \text{the 2D background RMS map}$. We mask pixels that are bright in multiple exposures – which we consider a real source – and perform binary opening and dilation on the result to smoothen the edges. We also flag "dead" pixels with unusually low values ($\leq 5 \times \text{the 2D background RMS map}$). 

This algorithm produces a bitmask for each individual image, indicating where cosmic rays and satellite trails are located. This mask is then used in the stack creation, so we can exclude the flagged pixels from the mean pixel value. Users of our individual images should be aware our algorithm does not pick up faint satellite trails - this is mainly a cosmetic concern, since faint trails are usually not picked up as sources during cataloging and appear weak in the mean-combined stacks.

For users interested in ``cleaned'' individual images without cosmic rays and satellite trails, we provide a separate mask that replaces these flagged pixels with values from the photometric infilling algorithm created and made available by \cite{saydjari}. 

\subsection{Stack Creation}
\label{sec:stacking}
In DR0, we used Auriga Imaging's RegiStar software (\url{https://aurigaimaging.com/}) to align and mean-combine our stacked images. However, the RegiStar stacking algorithm is proprietary and lacks some methodological transparency. For DR1, we choose to recreate our stacks with open-source scientific software, including \texttt{astrometry.net} to match our individual exposures, and the \texttt{ccdproc} Python package \citep{ccdproc} to mean-combine the aligned exposures.

\begin{figure}
    \centering
\includegraphics[width=1\linewidth]{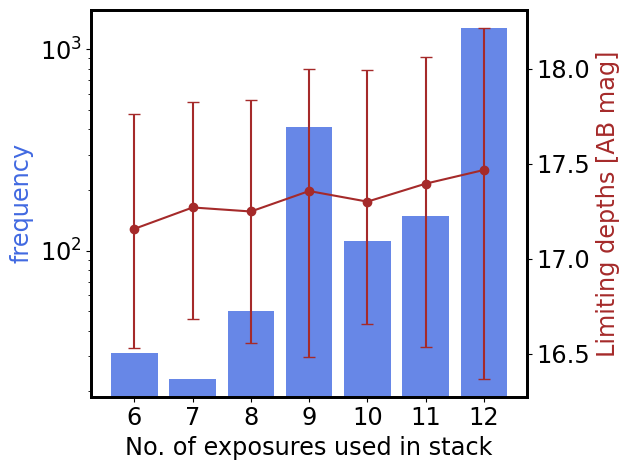}
    \caption{The frequency of fields (left axis, blue bars) and the range of their limiting depths (95th percentile of calibrated \ha magnitudes; right axis, red points) as functions of the number of  exposures used in the averaged stack. For their respective bins, the red dots represent the medians, and the 10th and 90th percentiles by the bars.}
    \label{fig:num_exposures}
\end{figure}

To avoid degrading the mean-combined stack quality, we do not use exposures that are flagged by either the background level or color-color offset outlier criteria outlined in Section \ref{sec:data_reduction}. Figure \ref{fig:num_exposures} illustrates the effect of this varying number of input exposures on the limiting depths of the final stack, defined as the 95th percentile of calibrated \ha magnitudes. We see that stacks made from fewer exposures (6-8) tend to reach shallower point source depths, with their median depths at \textasciitilde17.25 AB magnitudes, compared to those made from 9-12 exposures, which are deeper by \textasciitilde0.25 AB magnitudes.

As another method of quality control during the stacking process, we exclude using pixel values in the mean-combination if they are identified as cosmic rays or satellite trails from the stacking process (Section \ref{sec:cosmic_ray}). This reduces the risk of contaminating point-source photometry in the stack images, and subsequent cataloging. For these flagged cosmic-ray pixels, the final stack will evaluate the mean using non-flagged pixels from the remaining exposures.

During stacking, we first run \texttt{astrometry.net} with tuned parameters to minimize the variance of each exposure's astrometric solutions. This is crucial to avoid "blurriness" in the final stack, which can occur when sources are misaligned. While the final solutions of our released images use a distortion correction of order five, we generate intermediate solutions in the stacking process with an SIP correction of order three. This lower order avoids differential overfitting of the distortion solution for individual exposures for a given field, enabling a sharp final stack. 

The individual exposures are then reprojected onto the first exposure's coordinate frame, using the \texttt{reproject\_exact} function in the astropy API. This algorithm computes spherical polygon intersections, similar to the flux-conserving "drizzle" method developed by the Space Telescope Science Institute. The final stack is then trimmed to a $4000 \ \mathrm{pix}\times 4000 \ \mathrm{pix}$  frame to eliminate uneven borders, resulting from the minor pointing fluctuations between individual images (Section \ref{sec:stability}) which are made prominent after reprojection. Finally, both the stacks and individual exposures are solved with a polynomial distortion order of 5, as mentioned in Section \ref{sec:astrometry}, and all headers are updated to reflect this astrometric solution.

\subsection{Point source extraction \& matching}
\label{sec:extraction_matching}
\begin{figure}
    \centering
    \includegraphics[width=1\linewidth]{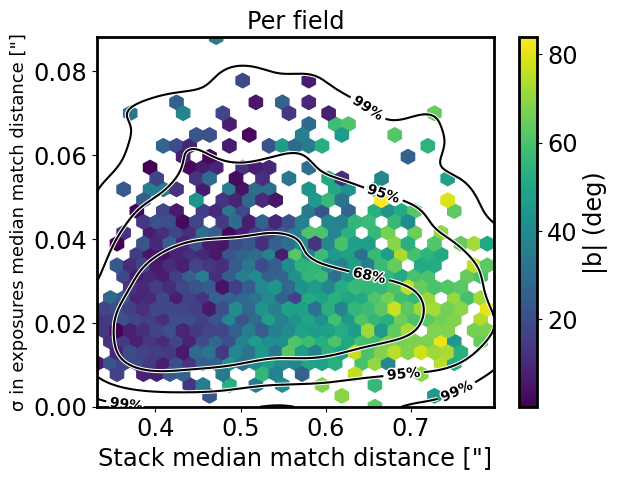}
    \caption{For each DR1 field, the median match distance to Pan-STARRS1 for the stack catalog compared to the the standard deviation of median match distance for the individual exposures. The sources are colored by the absolute value of their Galactic latitude. The match distances in the individual images have a subpixel deviation of at most 0.1\arcsec, and we see a decrease in match distance near the Galactic plane.}
    \label{fig:sep2d}
\end{figure}

To produce point-source catalogs for our images, we first extract the coordinates and fluxes of likely sources, and then match them to the Pan-STARRS1 DR1 mean catalog. Before extracting sources, we use \texttt{photutils} to estimate a background to subtract from the image. We use a median background estimator, with sigma-clipping done at $5\sigma$ over 5 iterations. We divide each image into boxes with length 16 pixels (box length), in which the background estimator is applied, and apply a 2D median filter with a window size of 3 pixels. If we overestimate our images to have a median PSF FWHM of 2.5 pix, a box length of 16 pix is sufficient to smooth away most sources. A filter length of 3 pix preserves much of the intricate background seen in the \ha wavelength, while allowing us to subtract a relatively unblemished background from our images for source extraction. A caveat of this method of background estimation is its limited performance in regions with prominent filamentary structure (e.g. the Galactic Plane), where flux estimates become more uncertain and require special consideration of the local background. In DR2, we expect to incorporate more complex photometric pipelines that can better estimate the flux (and flux error) for areas with structured background, such as the algorithm detailed by \cite{saydjari}.

To inform the parameters of our source extraction, we first calculate the median PSF FWHM of each image. We use the \texttt{DAOStarFinder} algorithm to find the 500 brightest, unsaturated stars (no pixels with count $>$ 60,000 ADU) with an initial, overestimated FWHM of 2.5 pix.  We then perform PSF photometry with a circular Gaussian point-response function (PRF) model to find the characteristic PSF FWHM value for the image. In Section \ref{sec:star_removal}, we discuss how the \texttt{astropy} functions for these processes can overestimate the FWHM value - this is not a concern in this case, since we find brighter sources are less prone to erroneous FWHM values.

To extract the full set of source coordinates, we run \texttt{DAOStarFinder} on the background-subtracted image and pass in the image's characteristic FWHM. We set a detection threshold of 4 $\times$ the median background root mean square (RMS) for our stack images, and 3 $\times$ for the individual images. We run photometry to measure source fluxes, as we do for the characteristic PSF. We ignore sources in the individual images if their centroid positions overlap with any cosmic rays or satellite trails, per Section \ref{sec:cosmic_ray}.

We then match these detections to the Pan-STARRS1 DR1 catalog (queried per Appendix \ref{appendix:cat_phot_supp}) \citep{ps_dr1_catalog}, with a 1.5\arcsec\ separation limit. When matching the Pan-STARRS1 catalog to the MDW stacks, we only consider Pan-STARRS1 objects with mean $r \leq 18$, and mean $r \leq 17$ when matching to our individual exposure catalogs. We incorporate IGAPS data in the source catalogs as a reference to a well-calibrated \ha point source survey. Since IGAPS has a  comparable pixel scale (0.33") to Pan-STARRS1, and both surveys base their astrometry on Gaia \citep{panstarrs_astrometry, IGAPS}, we match IGAPS sources directly to Pan-STARRS1 with a matching radius of 0.12". 

 Figure \ref{fig:sep2d} illustrates the median MDW-Pan-STARRS1 match distance (MMD) of each field's stack catalog, compared to the variation in MMDs across their individual catalogs ($\sigma_{\mathrm{MMD}}$). We color each bin by the median field Galactic latitude (|b|), and plot 3-$\sigma$ contour lines for the field density. For 95\% of fields, the $\sigma_{\mathrm{MMD}}$ for the individual catalog varies by less than 0.06\arcsec, indicating consistency among their astrometric solutions from Section \ref{sec:astrometry}. Figure \ref{fig:sep2d} also reveals a gradient pattern in stack catalog MMDs and their distance from the Galactic Plane, with fields closer to the Plane (center |b| $\leq 20^\circ$) having shorter MMDs of \textasciitilde0.4\arcsec, while the MMDs can exceed 0.7\arcsec\ for fields furthest from the Plane (center |b| $\geq 60^\circ$). This may be attributed to increased point-source crowding approaching the Plane, decreasing the distance for the matching algorithm to find a Pan-STARRS1 match. This may also be related to our astrometric error, discussed in Section \ref{sec:dp_catalogs}.
 
 For fields very close to the Plane (center |b| $\leq 10^\circ$), their stack catalog MMDs increase up to 0.6\arcsec. This is likely because we detect fewer sources in these fields, due to extinction in the Plane, raising the MMDs to be equivalent to that of fields at |b|\ $\approx 40^\circ$. The individual catalogs for some of these fields also vary more in their match distances, with $\sigma_{\mathrm{MMD}} \approx 0.07$\arcsec, which may be attributed to decreased point-source signal in regions abundant in diffuse gas.
 
\subsection{Point source flux calibration}
\label{sec:psm_flux_calibration}

\begin{figure*}
    \centering
\includegraphics[width=0.9\linewidth]{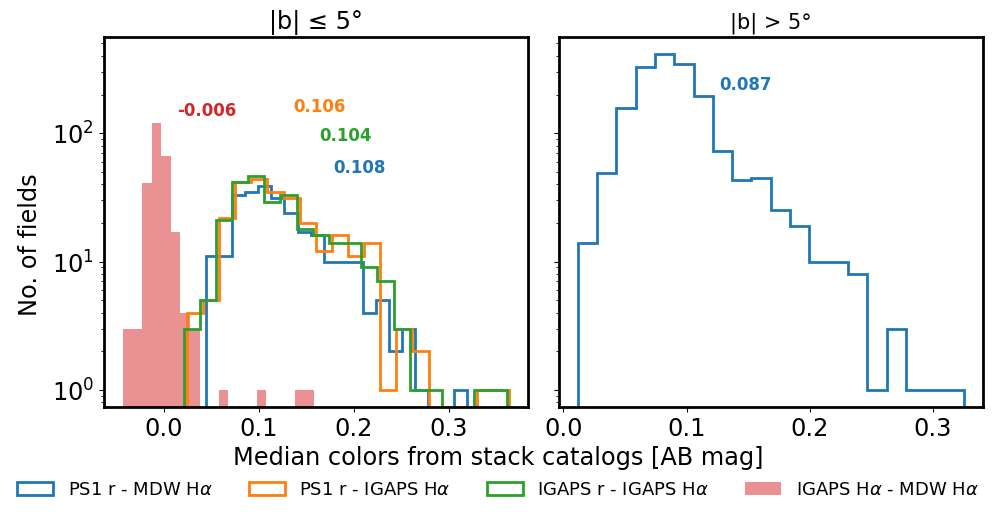}
    \caption{A comparison of the various colors relevant to the DR1 point-source \ha flux calibration, within and beyond the Galactic Plane (Pan-STARRS1 $r$ - MDW \ha in blue; Pan-STARRS1 $r$ - IGAPS \ha in orange; IGAPS $r$ - IGAPS \ha in green).  For $|b| \leq 5^\circ$ (left), we calibrate to IGAPS \ha, which gives us good agreement when comparing to broadband measurements from IGAPS (shaded red) and Pan-STARRS1. Outside the Plane (right), we calibrate to Pan-STARRS1 r, giving us close agreement.}
    \label{fig:stack_colors}
\end{figure*}

Since the MDW Survey does not image in broadband filters, we make use of our Pan-STARRS1 and IGAPS matches to calibrate the point-source photometric measurements in our stack and individual images. Beyond the Galactic Plane, we usethe $r$ magnitudes of our Pan-STARRS1 matches. Our strategy for these sources remains similar to the method described in \cite{MDW_DR0_paper}, where we find the stellar locus of the $r-\mathrm{H}\alpha$ vs. $r-i$ color-color plane and offset it to an expected value. However, instead of offsetting the  locus to a fixed target $r-\mathrm{H}\alpha$ for all fields, we now calculate the field-specific $r-i$ peak. We then offset our \ha magnitudes to match the corresponding $r-\mathrm{H}\alpha$ for this $r-i$ value, using the synthetic, unreddened main sequence tracks provided by the Modules for Experiments in Stellar Astrophysics (MESA) Isochrones and Stellar Tracks (MIST) software \citep{MIST_2, MIST_1}. Since most diffuse gas, which causes reddening through extinction, is clustered around the Galactic Plane, which also hosts older and inherently redder stars, it is reasonable to compare to the unreddened main sequence beyond the Plane. This is discussed in greater detail in Appendix \ref{appendix:cat_phot_supp} (Figure \ref{fig:ps_igaps}), where we compare the MIST tracks to the Pan-STARRS1 and IGAPS colors.

Reddening is a more significant concern when calibrating \ha fluxes in the Northern Galactic plane, which is surveyed by IGAPS (a footprint of $|b| \leq 5^\circ$). Here, we calibrate against the \ha  magnitudes from the matched IGAPS catalog for our stack and individual images. We find the peak of the IGAPS H$\alpha -$ MDW \ha distribution in each plane field, and offset our \ha magnitudes such that this difference at the locus is 0. We find it simpler to aim for a target difference of 0, since it can be tricky to calibrate against narrow-band \ha filters with different bandwidths (3nm in the MDW Survey, and 9.5nm in IGAPS). The observed \ha magnitudes of different spectral types can be sensitive to the filter width (discussed in Section \ref{sec:synphot}; also see \cite{VPHAS}). Per Section \ref{sec:variability}, DR1 detects a variety of spectral types, so a target difference of 0 allows for simplicity.

Figure \ref{fig:stack_colors} illustrates the distribution in our stack catalog of the relevant colors checked during the point-source calibration. Within the Galactic Plane, we see the median difference between MDW and IGAPS \ha is $-0.006$ AB magnitudes (red shaded region). The median difference between Pan-STARRS1 $r$ and MDW \ha (0.111; solid blue line) is only 2 millimagnitudes off from the median Pan-STARRS1 $r -$ IGAPS \ha (solid orange line) and 4 millimagnitudes off from IGAPS $r -$ \ha (solid green line). The distribution of the 3 median colors also share similar shapes, peaking at 0.1 magnitudes and dropping off to 0.3 magnitudes. Outside the Plane, where we calibrate to Pan-STARRS1 $r$, the median Pan-STARRS1 $r$ - MDW \ha (0.089) is nearly equivalent to the median Pan-STARRS1 $r -$ IGAPS \ha (0.106) and IGAPS $r -$ \ha (0.104) within the Plane. This agreement with Pan-STARRS1 and IGAPS suggests that our calibrated \ha magnitudes are generally reliable, both within and outside the dusty Milky Way plane, and supports our method of calibrating to broad-band/narrow-band magnitudes of external survey catalogs. Further discussion of the scatter in our photometric precision is given in Section \ref{sec:dp_catalogs_color_color}.


DR1 fields have varying levels of point-source crowding, since the dataset spans a large range of Galactic latitudes ($-60^\circ< b < 90^ \circ$). To find the peaks of the above-mentioned distributions in color-color space, while accounting for the wide range in source density across the plane (see Figure \ref{fig:color_color}) - which would skew more conventional measures like the median or mode - we use the 'auto' histogram binning strategy from the NumPy API \citep{numpy}. This strategy chooses the minimum bin size between the Sturges and Freedman–Diaconis estimators, providing reliable detection of the stellar locus regardless of the number of points in the underlying distribution. 

During calibration, we avoid using Pan-STARRS1 matches with  magnitudes either averaged from poor-quality detections, or not used in the Pan-STARRS1 relative photometry process. The specific checks are listed in Table \ref{appendix:ps1_flags}, and are included with the released catalogs in the \texttt{passes\_ps\_quality\_check} column.

\subsection{Star-removal \& Infilling}
\label{sec:star_removal}
In DR1, we improve both our star-removal algorithm and our infilling method to better facilitate the use of our star-removed images for studying diffuse and filamentary structure. In DR0, we flag pixels above a certain threshold using the image segmentation API from \texttt{photutils}, which required care to avoid flagging highly nebulous regions as "stars". This can occur because the image segmentation algorithm is attuned to detect both extended and point sources. In DR1, we instead use the detection catalog produced in Section \ref{sec:extraction_matching} - our source detection relies on the DAOPHOT algorithm, which only flags a "star" if it is well fit by a Gaussian point-spread function \citep{daofind}. This reduces the possibility of removing galaxies or high-nebulosity areas by mistake.

For optimal point source masking, we need to know the PSF FWHM.  We find that the \texttt{photutils} PSF photometry API can overestimate the fitted FWHM for some brighter sources, and we discuss our method for refitting the FWHM vs. magnitude function in Appendix \ref{appendix:fwhm_refitting} (Figure \ref{fig:fwhm_refitting}). To perform the masking consistently over all DR1 stacks, we perform an exponential function fit to source masking radius, $r$, vs. median FWHM, $f$, and magnitude, $m$ (Equation \ref{eqn:masking_radius}), so sources are appropriately masked based on their brightness.   To ensure the exponential function increases positively with brighter sources, we take the absolute value of their instrumental magnitude. The factor of $0.2$ is hand-tuned to prevent over-masking or under-masking sources. 
Finally, we use the infilling algorithm provided by \cite{saydjari} to replace the pixels of these masked stars with the statistical predictions of their mean pixel values according to the local background.

\begin{equation}
\label{eqn:masking_radius}
r(m, f) = 
\left( \frac{f}{2} \right) \times \exp(0.2 \times |m|) 
\end{equation}

 Our infilling does not perform well for highly saturated sources e.g. as observed in fields 60, 375, and 438. These sources require particularly large masks, which the DR1 pipeline's infilling parameters are not optimized to handle. As a result, the infilled pixels do not uniformly blend into the local background, like the infills of unsaturated sources do. For DR2, we will improve the parameters passed to the infilling API \citep{saydjari}, to better handle these cases. An example using the DR1 star-removed images for filament analyses is presented in Section \ref{sec:filaments}.


\section{Quality Analyses}
\label{sec:reshoots}
As we iterate through the pipeline described in Section \ref{sec:pipeline}, we perform a number of quality checks to flag fields that would benefit from additional reshoots. These checks are described below.

\subsection{Background levels}
\label{sec:bkg_levels_qa}

In general, low background levels are desirable and indicative of optimal observing conditions. In the initial imaging stages, data are typically acquired in groups of 3 exposures, each being 20 minutes within 1-hour blocks. If one exposure is adversely affected, it is likely that the other two in the group are similarly compromised. This contributes to the majority of reshoot selection in the dataset. Rather than adopting a fixed background threshold for triggering reshoots—which might fail to account for natural spatial variations in background brightness—we define a field-specific baseline using the average of the six lowest median background values available for that field.

In cases where this baseline is significantly higher than those of neighboring fields, all exposures are manually inspected to determine whether the elevated background arises from true astrophysical variation or from image degradation that warrants reshooting. For routine evaluation, any exposure with a median background exceeding the field-specific baseline by more than 140 counts—a threshold derived from empirical comparisons with rejected images—is flagged as suboptimal. Fields meeting this criterion are then prioritized for reshoot scheduling.

\begin{figure}
    \centering
    \includegraphics[width=1\linewidth]{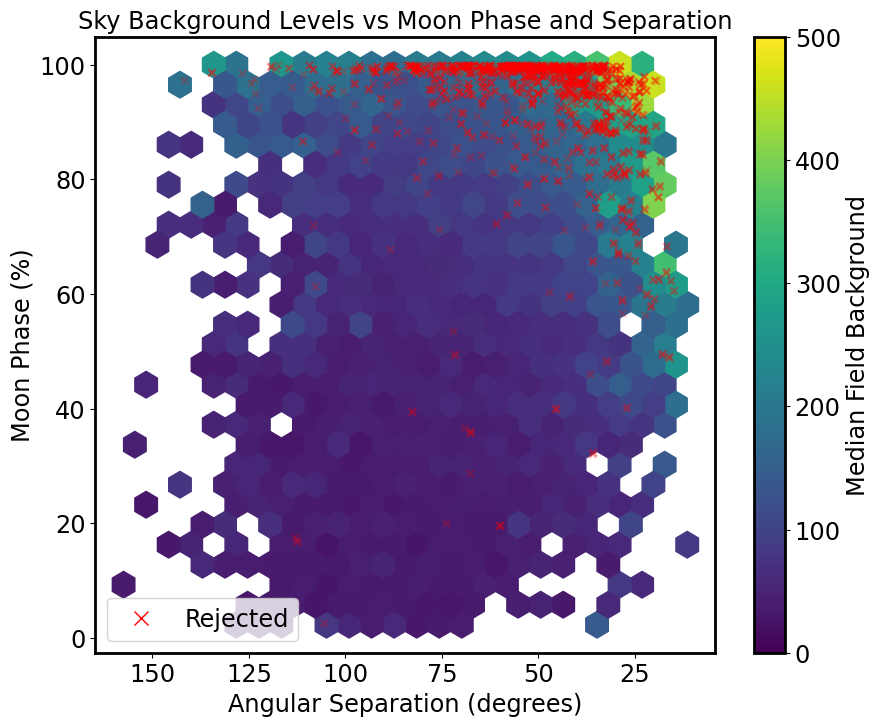}
    \caption{Median Background of original individual exposures plotted against the moon separation angle and moon phase at the time of observation. Individual images considered sub-optimal are flagged with red crosses, which is more likely to happen with inadequate moon avoidance combined with a full moon. The newer images and reshoots have strict moon avoidance protocols implemented.}
    \label{fig:moon_phase}
\end{figure}

Through retrospective analysis, we identified that one of the primary sources of elevated background levels in flagged exposures is insufficient moon avoidance in scheduling constraints. By cross-referencing exposure background levels with lunar phase and angular separation from the Moon, we observed a clear correlation between lunar proximity and contamination, as illustrated in Figure \ref{fig:moon_phase}. Exposures identified as suboptimal based on the threshold criteria are marked by the red crosses.

In addition to lunar interference, other sources of irregular background contamination include the presence of sand or dust, unwanted local light sources, and minor operational inconsistencies. These cases, while less common, are detected through targeted inspection and contribute to background variability that may necessitate additional quality control measures.

\subsection{Flux calibration offsets}
\label{sec:flux_calibration_qa}

\begin{figure}
    \centering
    \includegraphics[width=1\linewidth]{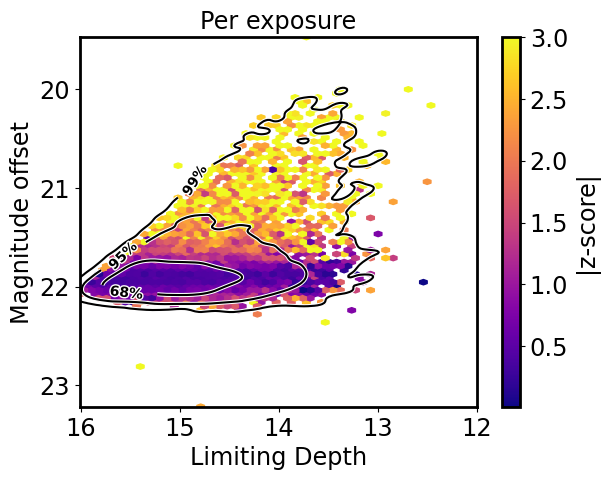}
    \caption{For each exposure, the z-score of its point-source calibration offset when compared to their sibling exposures. Images that deviate further from the field's norm (higher z-score) tend to require larger offsets and reach shallower depths.}
    \label{fig:flux_offset_z}
\end{figure}

After determining the calibration offset of our images in Section \ref{sec:psm_flux_calibration}, we compare the values of each exposure to find outliers that may not have been flagged from the background criteria. Brief inclement weather phenomena, like passing clouds or dusty conditions, would block light from reaching an MDW telescope as it observes a field, resulting in an image with relatively low signal. These images can be singled out by the relatively large calibration offset needed, since their recorded instrumental fluxes are further from the "true" flux. For each field, we sigma clip the \textasciitilde12 offsets at 3$\sigma$ over 5 iterations, evaluate the clipped median and standard deviation, and use it to calculate a z-score for each image. Figure \ref{fig:flux_offset_z} compares these scores to the total magnitude offset calculated for the exposure and their limiting depth. We see that, compared to their companion exposures, images with $|z| \geq 2$ tend to be shallower and their point-source instrumental magnitudes fainter. We flag exposures for reshooting if they have $|z| \geq 3$, eliminating the most egregious offenders.

\subsection{Checks for calibration frames}

It is crucial to ensure the integrity of calibration frames in order to minimize the propagation of systematic errors through the data reduction pipeline. Routine quality assurance checks are implemented at multiple stages to detect issues early in the workflow.

Common problems include prematurely terminated exposures due to misconfigured scheduling parameters or environmental interferences (e.g., dome or roof movements conflicting with exposure timing). Hardware malfunctions, such as with the flip-flats (automated lens cap that serves as a consistent flat-field illuminator for automated operation) are infrequent, but can occur during long-term survey operations and would affect the reliability of calibration data.

Initial quality checks are performed by parsing FITS metadata to confirm that exposure times match expected values and that instrumental parameters—such as CCD temperature and environmental factors—remain within nominal operating ranges. Following header validation, basic statistical metrics (mean, median, standard deviation) are computed to identify outliers or anomalies suggestive of non-uniform illumination or other defects. Frames failing these checks are excluded from further processing.

Visual inspection remains an important complementary method, particularly for identifying artifacts such as light leaks, vignetting, or cosmetic defects. Image snapshots are reviewed to flag visually distorted or contaminated frames.

In rare instances, calibration frame issues are not immediately obvious and only become apparent in the stacked science images, where subtle background inconsistencies or pattern noise are amplified. In such cases, we trace the problem back to the original calibration set, reselect appropriate frames, and recalibrate the affected science exposures before regenerating the stack. This correction process ensures the final data products meet the quality standards and consistencies.

\section{Data Products}
\label{sec:data_products}

\begin{table}[]
\begin{tabular}{|c|c|c|}
\hline
                                   & \textbf{Product Type}                 & \textbf{Data Products}           \\ \hline
\multirow{7}{*}{\textbf{Images}}   & \multirow{3}{*}{Mean-combined stack \textsuperscript{a}}  & Instrumental                     \\ \cline{3-3} 
                                   &                                       & Calibrated                       \\ \cline{3-3} 
                                   &                                       & Star-infilled \textsuperscript{a}                    \\ \cline{2-3} 
                                   & \multirow{3}{*}{Individual exposures} & Instrumental                     \\ \cline{3-3} 
                                   &                                       & Calibrated                       \\ \cline{3-3} 
                                   &                                       & Cosmic-ray infill \textsuperscript{a}               \\ \cline{2-3} 
\multirow{2}{*}{\textbf{Catalogs}} & Mean-combined stack                   & \multirow{2}{*}{Matched catalog \textsuperscript{b} } \\ \cline{2-2}
                                   & Individual exposures                  &                                  \\ \hline
\end{tabular}
\begin{tablenotes}
    \small 
    \item{$^\text{a}$ \footnotesize Each type of stack image is cleaned of cosmic-rays and satellite trails, per Section \ref{sec:cosmic_ray}.}  
    \item{$^\text{b}$ \footnotesize The bitmasks for infilled/flagged pixels are also provided.}    
    \item{$^\text{c}$ \footnotesize Catalogs are matched to the Pan-STARRS1 DR1 survey, with a subset of sources in the Northern Galactic Plane matched to the IGAPS survey.}    
\end{tablenotes}
\caption{Reference table for the released data products in DR1.}
\label{tab:data_products}
\end{table}

The final DR1 dataset covers 2116 fields, with each image being roughly 60 MB (totaling \textasciitilde2 TB), and 
about 30 GB of catalogs (compressed using gzip). All DR1 data products can be accessed at the MDW Survey's website: \textbf{\url{https://mdw.astro.columbia.edu/}}. The DR1 catalogs are additionally available in the AAS Journals Zenodo repository:\dataset[doi:10.5281/zenodo.17307324]{https://doi.org/10.5281/zenodo.17307324}. Since the DR1 dataset is several times larger than the preceding DR0, the components of the release are listed in Table \ref{tab:data_products}.

\subsection{Point source catalogs}
\label{sec:dp_catalogs}
We release two sets of point source catalogs: a single catalog generated from all DR1 mean-combined stacks (50 million sources) and catalogs of each field's individual exposures. These catalogs consist only of sources matched to the Pan-STARRS1 DR1 catalog, of which a subset is also matched to IGAPS (Section \ref{sec:extraction_matching}). Appendix \ref{appendix:catalog_columns} outlines the catalog columns.

\begin{figure}
    \centering
    \includegraphics[width=1\linewidth]{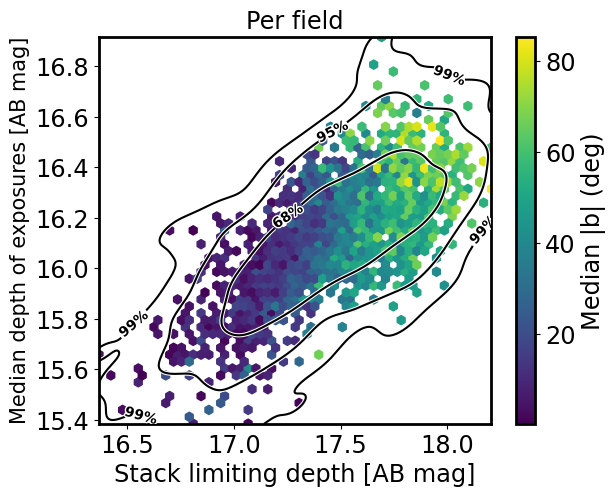}
    \caption{For each field, the limiting depth of the stack catalog (95th percentile of \ha magnitudes; see Section \ref{sec:dp_catalogs}) against the median depth of the individual images. The colorbar shows the absolute value of Galactic Latitude for each field. Both the individual and stack catalogs show a dependence on Galactic Latitude.}
    \label{fig:limiting_depths}
\end{figure}

For each field's catalogs  of interest (stack and individual images), we evaluate the 95th percentile value of the calibrated \ha magnitudes as a heuristic for the limiting depth, and display these values in Figure \ref{fig:limiting_depths}. Catalogs from our stacked images, exposed for a total of \textasciitilde4 hours, can reach depths of 16.5-18.0 AB magnitudes. The median of the individual images, exposed for 20 minutes, ranges from 15.4-16.8 AB magnitudes. The colorbar in Figure \ref{fig:limiting_depths} also illustrates the limiting depths as a function of distance from the Galactic Plane. Both the stack and individual images reach their shallowest depths when approaching the Plane (shaded dark blue; $|b| \leq 10^\circ$). The Plane stack images report depths of 16.5-17.25 AB mags, while the individual images can have wider ranges of 15.4-16.4. Since the Galactic Plane has high column densities of \ha-bright diffuse gas, the individual images here are dominated by the background noise, which limits the depth to which we can detect sources even after mean-combination. In Figure \ref{fig:limiting_depths}, fields far from the Plane ($|b| \geq 60^\circ$) are shaded bright green and yellow. Here, the individual catalogs can reach their shallowest depths (16.2-16.8 AB mags), while the stacks perform relatively better at depths of 17.5-18 AB mags. Since the Milky Way's halo contains fewer stars and lower gas column densities compared to the Galactic Plane, stacking our exposures improves the field signal-to-noise ratio.

\begin{figure}
     \centering
     \includegraphics[width=1\linewidth]{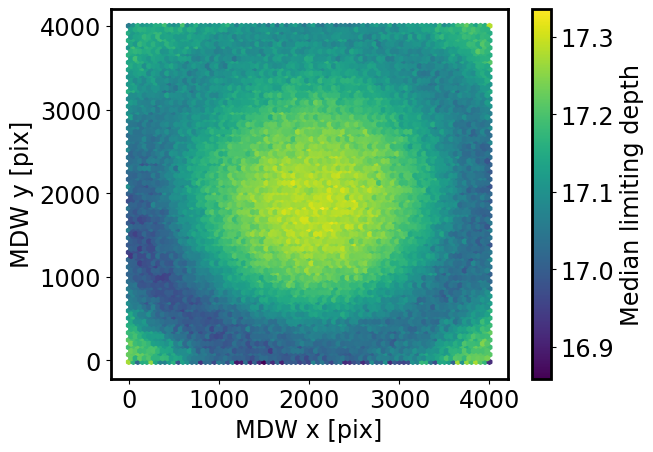}
     \caption{ Across all DR1 matched sources, the median limiting depth (95th percentile of \ha magnitudes) as a function of its (x, y) coordinate. This reveals a radial pattern, caused by field-dependent variation in detector sensitivity, and is discussed in Section \ref{sec:dp_catalogs_color_color}.}
     \label{fig:mdw_ha_2dhist}
 \end{figure}


Taking the DR1 stack catalog as a whole, we also notice a spatial pattern in \ha magnitudes.
Figure \ref{fig:mdw_ha_2dhist} displays the x and y coordinates for all DR1 matches, with bins colored by the median limiting depth, defined as the 95th percentile of calibrated \ha magnitudes. This reveals a ring-like pattern, where we seem to be able to detect fainter sources closer to the field center, while the median detection for sources \textasciitilde1500 pixels from the image center tend to be brighter by roughly 0.5 AB magnitudes. This dropoff is likely caused by variations in detector sensitivity.

 \begin{figure}
     \centering
     \includegraphics[width=1\linewidth]{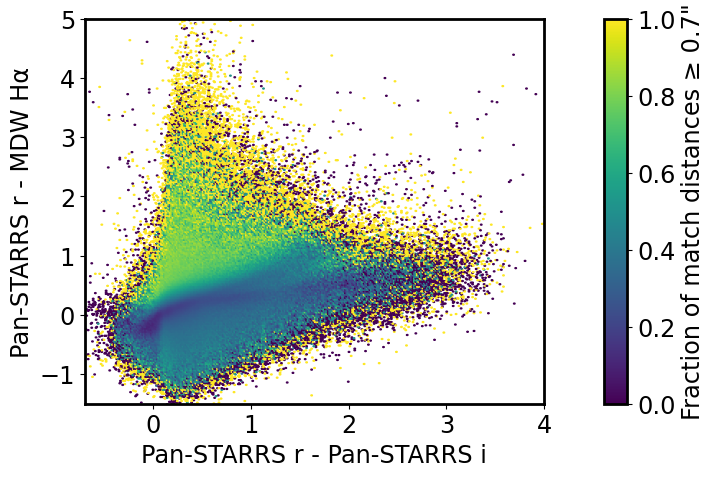}
     \caption{For all DR1 matches with quality Pan-STARRS1 information, their $r-\mathrm{H}\alpha$ color against $r-i$. The bins are shaded by the fraction of Pan-STARRS1 matches with separation distance greater than or equal to 0.7". Discussion can be found in Section \ref{sec:dp_catalogs}. }
     \label{fig:match_fractions}
 \end{figure}

While our astrometry has evolved from the DR0 dataset (Section \ref{sec:astrometry}), we find there is still room for improvement with regard to our astrometric error. Figure \ref{fig:match_fractions} shows the $r-\mathrm{H}\alpha$ vs. $r-i$ color-color plane for all DR1 stack matches, with each bin colored by the fraction of sources with match distance greater than or equal to 0.7\arcsec (our maximum limit for matching is 1.5\arcsec). We see that the fraction is less than 50\% (medium blue) for most bins and falls to lower than 30\% (dark blue) for sources neatly along the main sequence (see Figure \ref{fig:color_color} for reference). However, there is a stark difference for sources above the main sequence tracks, where at least 80\% (bright green) of binned sources have match distances above 0.7\arcsec. There are also numerous bins shaded at 100\% (yellow) bordering the color-color plot, signaling likely mismatches. However, it is worth noting the scattering of bins shaded at 0\% (darkest blue), where sources have quite good match distances to Pan-STARRS1 - so not all of these outliers are due to poor matching. We intend to investigate this further for DR2. The color-color plots, filtered to DR1 sources with good match distances and signal-to-noise, are detailed in Section \ref{sec:dp_catalogs_color_color}, and the color-color outliers are touched on in Section \ref{sec:variability}.

 

\subsubsection{Color-color planes: $r - \mathrm{H}\alpha$ against $r - i$, $g - r$}
\label{sec:dp_catalogs_color_color}

By displaying sources in a color-color plane, we can evaluate the effects of reddening and estimate the spectral types of point sources in the DR1 catalog (further investigated in Section \ref{sec:variability}). To do so, we use the g, r, and i magnitudes of the Pan-STARRS1 matches with our calibrated \ha magnitudes. We consider $r - \mathrm{H}\alpha$ against two colors: $r-i$ to examine the effect of reddening and $g-r$ to examine the upturn of cooler K-M stars. From the MIST library, we extract synthetic photometry of the Pan-STARRS1 g/r/i and the IPHAS \ha filters (note that the IGAPS survey encompasses IPHAS) for the unreddened main sequence (also see Appendix \ref{appendix:ps1_flags}). We also use the bolometric corrections provided by MIST ($0 \leq A_v \leq 6$, $R_v=3.1$) to generate reddening vectors for the A2V, F5V, and M0V spectral types. Finally, we use the following catalog checks, per discussion in Section \ref{sec:dp_catalogs}, to refine our sample to likely Pan-STARRS1 matches with quality photometric data (resulting in 20 million sources, out of the total 50 million): 

\begin{lstlisting}[language=Python, breaklines=true]
quality_ps1_information = catalog['passes_ps_quality_check']
likely_match = (catalog['n_ps_nearby'] == 1) & (catalog['ps_sep2d'] <= 0.7)
good_signal_to_noise = catalog['MDW_Ha_mag'] <= 16.5
catalog_subset = quality_ps1_information & likely_match & good_signal_to_noise
\end{lstlisting}
 
\begin{figure*}
    \centering
    \includegraphics[width=1\linewidth]{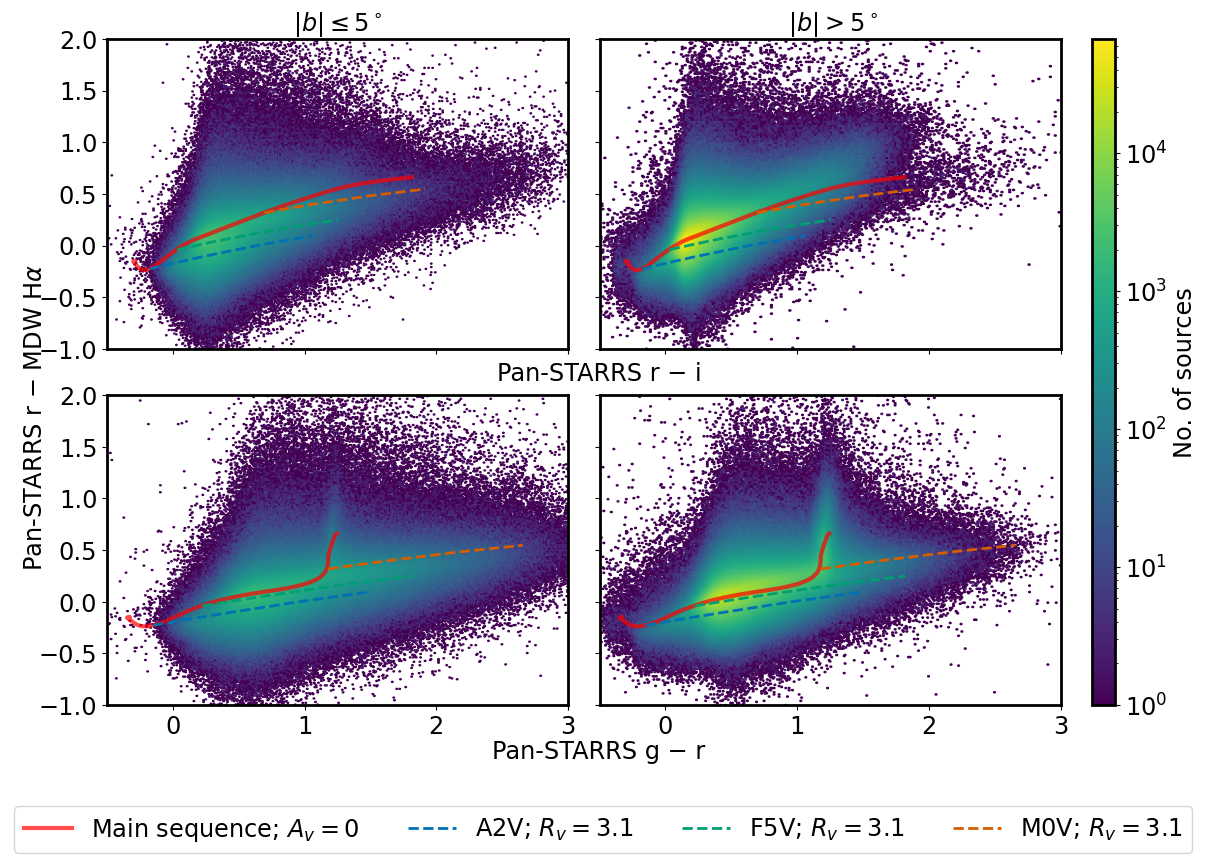}
    \caption{For the matched stack catalog, the $r-\mathrm{H}\alpha$ color against $r-i$ (top) and $g-r$ (bottom), for sources within the Plane (left) and beyond (right). We overlay synthetic tracks from MIST, for the main sequence (solid red) and the reddening vectors (up to $A_v=6$) for A2V, F5V, and M0V spectral types (blue, green, and orange dashed lines, respectively). The effect of reddening on our sources is significant in the Plane, and less so outside it. Further discussion can be found in Section \ref{sec:dp_catalogs_color_color}.}
    \label{fig:color_color}
\end{figure*}

 These results are displayed in Figure \ref{fig:color_color}, separated by populations within and beyond the Galactic Plane. For sources within the Plane, we see the clear effects of extinction and reddening. We find a long tail of sources on the redder end of $r-i$ ($\geq 2$) and $g-r$ ($\geq 2.5$), with very few following the synthetic unreddened main sequence (solid red line) - the density of sources in this area of the plot is low enough that we cannot properly distinguish the unreddened main sequence feature. The faint upturned spur in the $g-r$ plot helps reveal some unreddened K-M stars (\textasciitilde$10^4$ sources per bin). Otherwise, sources generally follow the reddening vectors for various spectral types, plotted in blue, green, and red dashed lines (A2V, F5V, and M0V, respectively). This reddening effect is caused by the presence of dust and gas within the Plane, and is likely compounded by the absence of continuum subtraction in our photometry. 
 
 Conversely, outside the Galactic Plane, we see significantly more sources following the unreddened main sequence tracks, and far fewer reddened sources. Of note is the prominent, upturned spur at $g-r \approx 1.25$, corresponding to cool K and M stars. This feature appears to extend further in our source catalog, up to $r-\mathrm{H}\alpha \approx 1.5$,  than in IGAPS, which extends only  to a maximum of $r-\mathrm{H}\alpha \approx$ 1 (left of Figure \ref{fig:ps_igaps} in Appendix \ref{appendix:cat_phot_supp}). 
 
 Looking at the $r-i$ color (top right in Figure \ref{fig:color_color}), we also see a diffuse feature for sources at $r-i \approx 1$, aligning with where K-M types theoretically reside. The presentation of these spectral types in the color-color plane (synthetic and observed) is discussed in \cite{VPHAS}, and is thought to be sensitive to filter bandwidth. We touch on this in Section \ref{sec:synphot}, where we see brighter expected \ha magnitudes in the narrower MDW filter than in IGAPS, for synthetic K-M spectral types. In Figure \ref{fig:color_color}, we may be seeing the observed effect of our narrower filter. Observing brighter \ha magnitudes for K-M type stars would explain why our vertical $g-r$ feature extends further than sources in the IGAPS catalog, because our $r-\mathrm{H}\alpha_{\mathrm{MDW}}$ values mathematically become larger than for $r-\mathrm{H}\alpha_{\mathrm{IGAPS}}$. This same reasoning can explain why our sources on the redder end of $r-i$ ($\geq$ 1) present in a more diffuse feature, fanning out over a range of $\approx 0.5$ magnitudes, than in IGAPS ($\approx 0.25$ magnitudes).   

 The top right plot in Figure \ref{fig:color_color} also reveals a prominent vertical feature at $r-i \approx 0.25$ magnitudes, with $r-\mathrm{H}\alpha \approx [-1, 2]$. We believe this spread is partly explained by the presence of \ha-excess sources, as described in Section \ref{sec:variability}. While the cuts above refine our working catalog to the most likely Pan-STARRS1 matches, we believe this vertical feature is also partially explained by catalog mismatches for faint detections in dense fields or due to systematic astrometric errors. This is discussed above, in Section \ref{sec:dp_catalogs}.  



\begin{figure}
    \centering
    \includegraphics[width=1\linewidth]{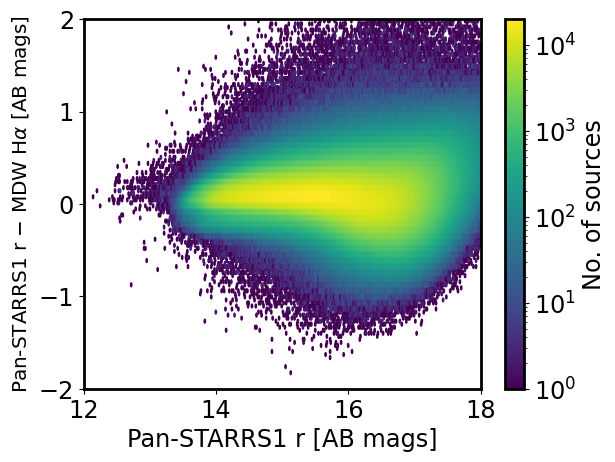}
    \caption{For our matched stack catalog (limited to Pan-STARRS1 sources with $r \leq 18$), the difference between Pan-STARRS1 $r$ and MDW \ha magnitudes as a function of $r$ magnitude. The bins are colored by source density. }
    \label{fig:rHa_r}
\end{figure}

 We can also more thoroughly investigate the scatter in our photometric precision (rather than just the centers, as discussed in Section \ref{sec:psm_flux_calibration}). Figure \ref{fig:rHa_r} illustrates this relative to our Pan-STARRS1 matches, where the stack catalog is filtered according to the example Python query above (except for the condition on brightness, so we can look across all magnitude ranges). The figure cuts off at $r = 18$ AB mag because that is the brightness limit imposed when matching to Pan-STARRS1 (Section \ref{sec:extraction_matching}). We see our Pan-STARRS1 $r$ - MDW \ha colors center at roughly 0.1 for most sources with $r \leq 16$ AB mag, and beyond this, the scatter in color can widen by up to $\pm 1$ AB mag. Regardless of Pan-STARRS1 $r$ magnitude, we see a relatively consistent spread in $r-\mathrm{H}\alpha$, in the positive direction - this is likely related to the discussion earlier in this section on astrometry accuracy and \ha-excess sources.

\subsection{Images}
\label{sec:dp_images}

All images include WCS information in the header by \texttt{astrometry.net}, as described in Section \ref{sec:astrometry}. For the individual images (Section \ref{sec:cosmic_ray}) and star-removed stacks (Section \ref{sec:star_removal}), we include the single-detection and source boolean masks as additional data units in the FITS file, respectively, so users can conveniently check which pixels have been modified.

We use the same methods described in Section 4 of \cite{MDW_DR0_paper} to convert the flux calibration offset from Section \ref{sec:psm_flux_calibration} into an AB magnitude. This magnitude offset can be found in the FITS headers of the instrumental images, under the \texttt{MAGZP} keyword. This offset is subsequently also applied to the instrumental data to create flux-calibrated images, with pixel units in Jy, such that calibrated fluxes can be directly determined during source photometry.

\begin{figure}
    \centering
    \includegraphics[width=1\linewidth]{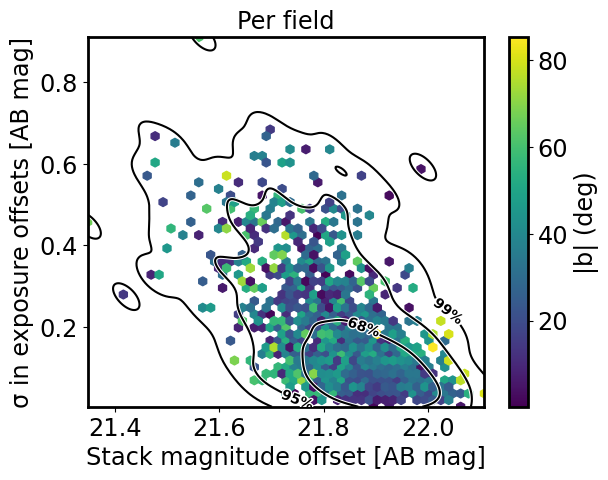}
    \caption{For each field, the calibration offset calculated for the stack images versus the standard deviation of this offset for the individual exposures (calculated in Section \ref{sec:psm_flux_calibration}). Data points are colored by the field's Galactic latitude, revealing no obvious calibration biases, with further discussion in Section \ref{sec:dp_images}.}
    \label{fig:mag_offset}
\end{figure}

We find the final flux offset of the stacked images is correlated to the offsets determined for the input individual images. In Figure \ref{fig:mag_offset}, we see that a greater deviation in flux offset ($\sigma \geq 0.3$) among a field's exposures also leads to a larger offset in the mean-combined stack (offset $\lesssim 21.8$). This deviation is likely caused by temporal weather changes (e.g. those outlined in Section \ref{sec:flux_calibration_qa}) that prevent light from reaching our telescopes and require a greater calibration offset for the affected exposure. However, for most fields, the calibration offset in the exposures varies by less than 0.2 AB magnitudes, suggesting most of our images are of consistent quality. Figure \ref{fig:mag_offset} also reveals no discernible pattern with Galactic latitude, assuring us that our point-source calibration process is impartial to the effects of the dusty Plane and the sparse Halo.

\begin{figure}
    \centering
    \includegraphics[width=1\linewidth]{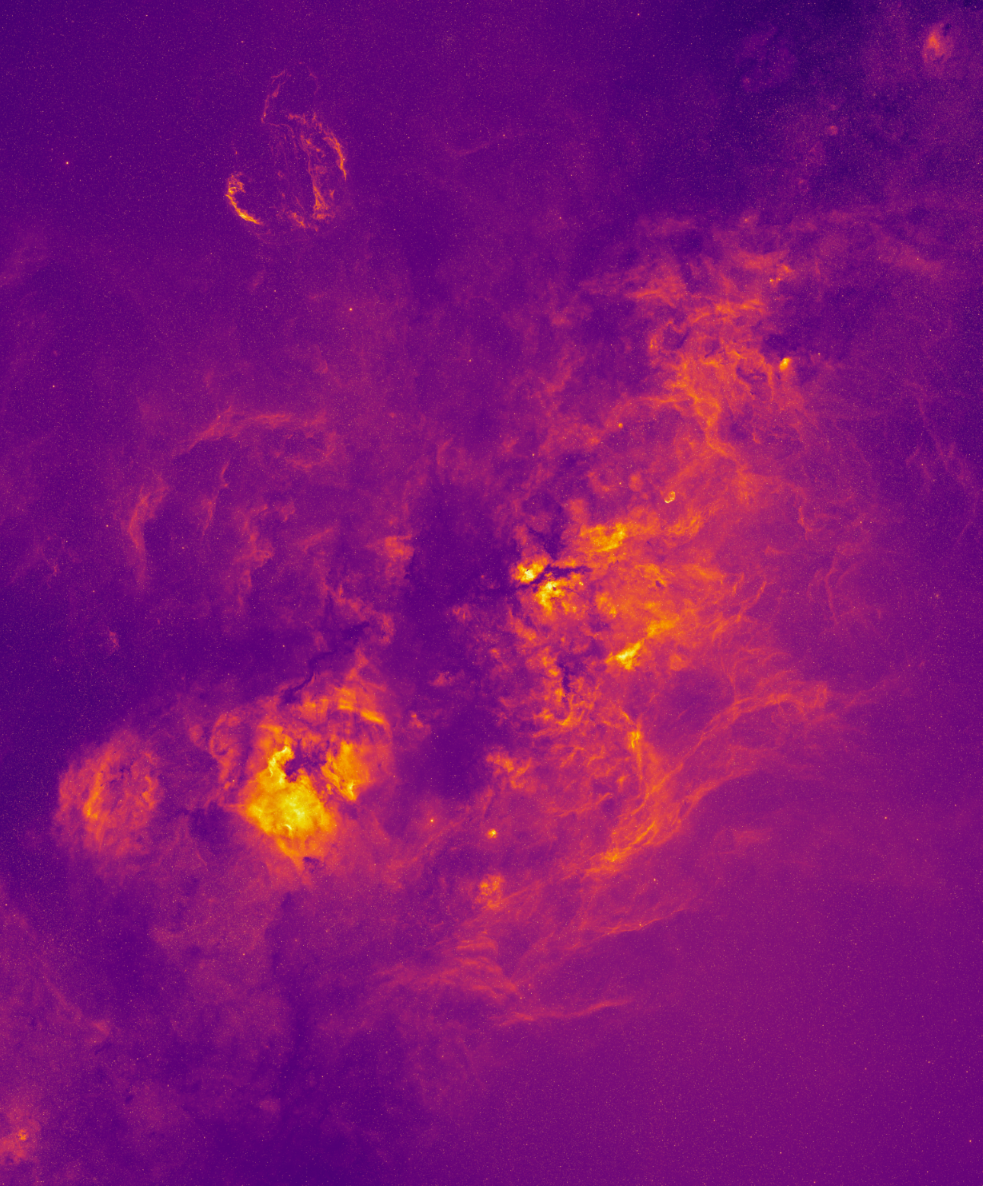}
    \caption{Mosaic of the Cygnus region, centered at $\mathrm{RA} = 311.5^{\circ}$, $\mathrm{Dec} = 39.0^{\circ}$ and of size $30.0^{\circ}$ by $24.0^{\circ}$. This image has been smoothed and the contrast scaling has been manually adjusted to bring out the diffuse features.}
    \label{fig:cygnus}
\end{figure}

We also use the same algorithm for mosaic images as discussed in Section 4.3 of \cite{MDW_DR0_paper}. We are able to improve the calculation efficiency due to improved stacked image qualities and more accurate astrometric solutions. We also trim the selected field images to assist the alignment during the process to minimize the overlap seen due to field background variations. Our default mosaic sizes are 10 degrees by 10 degrees, which in most cases captures the larger diffuse structures while being reasonably light on computation requirements. In specific regions we make mosaics with different shapes and sizes to capture known features in the area. For example, a mosaic of the Cygnus region is shown in Figure \ref{fig:cygnus}.

\subsubsection{Error Analysis}

We characterized the electronic gain and the read noise of the detectors using flat-field and bias frames over the period of the entire survey. These analyses were performed independently for each of the three telescopes to identify potential variations and long-term trends.

For each date with available data, we sampled all possible pairs of individual flat frames. Only the central regions of these images are used to minimize the effects of edge artifacts and vignetting. The gain was calculated for each pair by computing the ratio between the mean signals for both frames ($\mu_1$, $\mu_2$) and the variance of the pixel-wise difference between the two images ($F_1 - F_2$):
\begin{equation}
\text{Gain} = \frac{\mu_1 + \mu_2}{\mathrm{Var}(F_1 - F_2)}
\label{eq:gain}
\end{equation}

Read noise values were computed using a similar approach. In this case, we used pairs of bias frames, which contain no photo-electron signal and capture only the electronic noise from the readout process. Each pair was differenced to remove fixed pattern noise and isolate the stochastic component of the readout noise. The standard deviation of the resulting difference frame was then scaled by inverse of $\sqrt{2}$ and multiplied by the gain factor to obtain the read noise in electrons:
\begin{equation}
\text{Read Noise} = \frac{\sigma_{\mathrm{bias1} - \mathrm{bias2}}}{\sqrt{2}} \times \text{Gain}
\label{eq:readnoise}
\end{equation}

\begin{figure}
    \centering
    \includegraphics[width=1\linewidth]{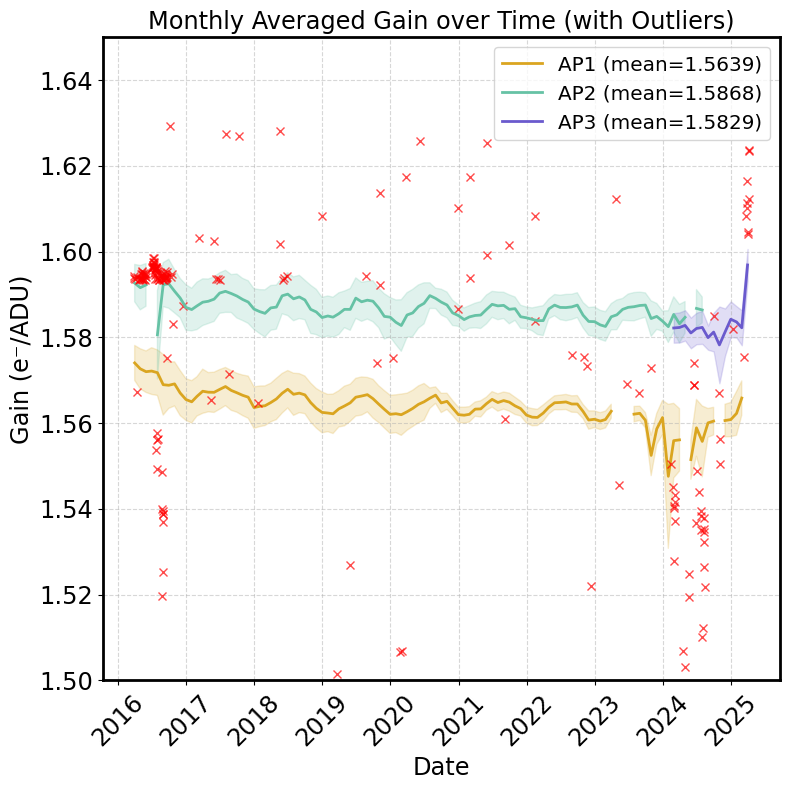}
    \caption{Gain factor over time for each of the 3 telescopes. The outliers are marked in red crosses, and are discarded from data processing. Because flat fields are taken very frequently, we show the binned monthly averages in this plot. }
    \label{fig:gainfactor}
\end{figure}

\begin{figure}
    \centering
    \includegraphics[width=1\linewidth]{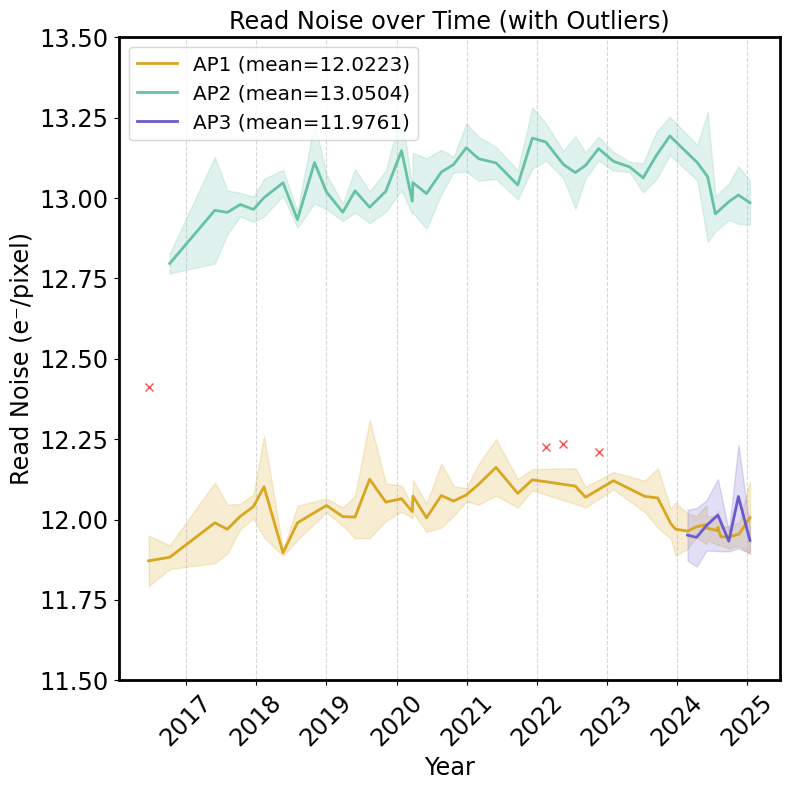}
    \caption{Read noise over time for each of the 3 telescopes. The outliers are marked in red crosses. Unlike flat frames, bias frames are not taken daily. To ensure timely correction, defective images are often identified immediately, thus making outliers rare.}
    \label{fig:readnoise}
\end{figure}

For both gain and read noise, nightly averages and standard deviations were computed by aggregating the results from all valid frame pairs. Figure \ref{fig:gainfactor} and \ref{fig:readnoise} shows the gain and read noise values over the full duration of the survey. The gain factors remained stable across time and among the three telescopes, with values of 1.56, 1.59, and 1.58~$e^-$/ADU, respectively. The gain factor shows small periodic variations annually, which we believe is due to the seasonal temperature change on parts of the optical system beyond the strictly thermal-regulated parts such as the CCD. The read noise showed slightly more variation between telescopes, with AP1 and AP3 exhibiting values near 12.0~$e^-$, while AP2 consistently displayed slightly elevated values around 13.1~$e^-$.

\section{Initial science}
\label{sec:initial_science}
There are a multitude of science uses of the MDW data set.   The data can be used to study HII regions, planetary nebulae, stars, and even nearby galaxies.   Here we describe two preliminary science results on ionized filaments in the direction of Lyra and stellar \ha variability and excess sources.  

\subsection{Filamentary structure}
\label{sec:filaments}
\begin{figure}
   \centering
   \includegraphics[width=1\linewidth]{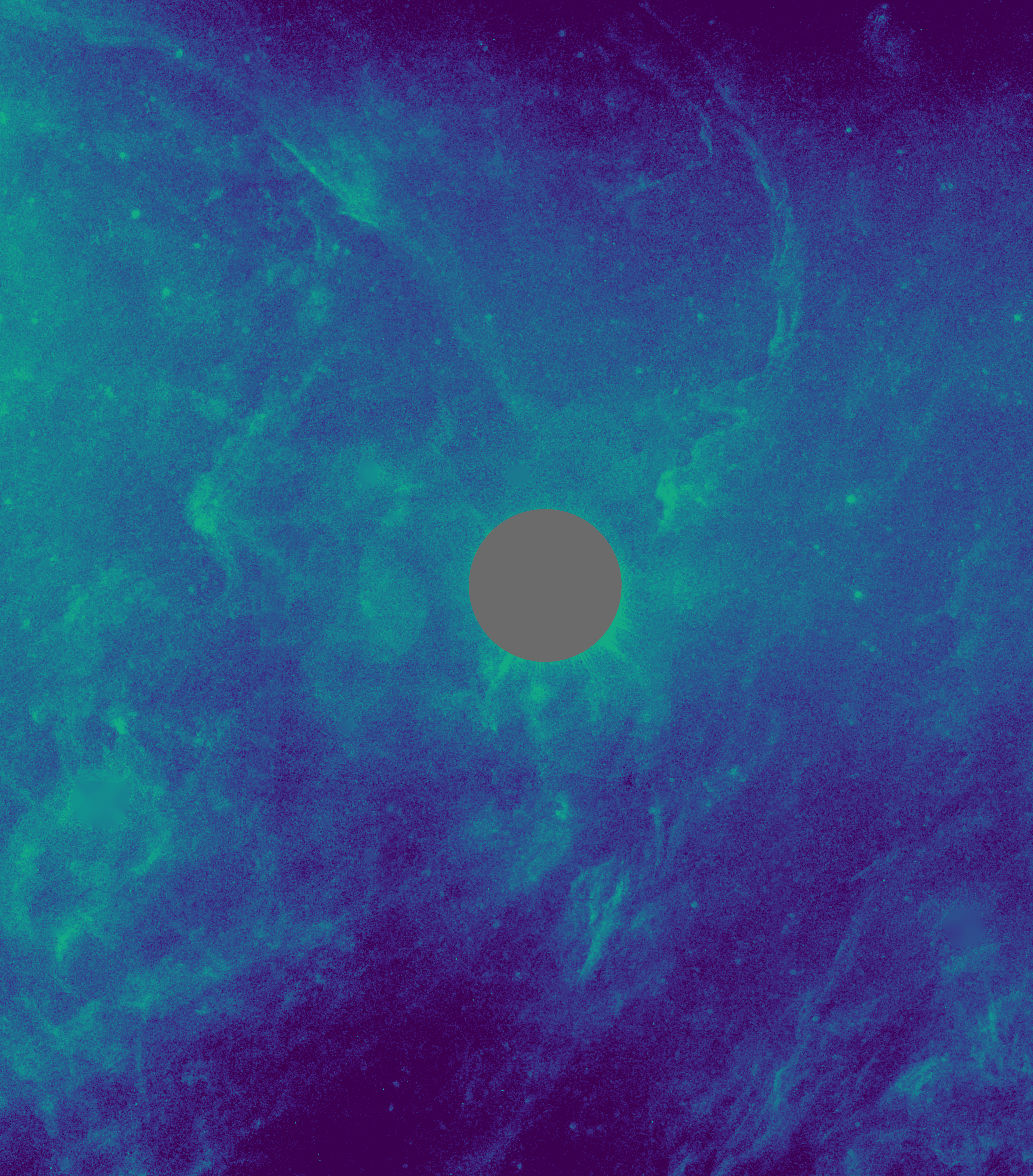}
    \caption{We show a region of size $9.9^{\circ}$ by $9.9^{\circ}$ centered at $\mathrm{RA} = 280^{\circ}$, $\mathrm{Dec} = 38.7^{\circ}$ in the direction of Lyra and the star Vega (masked). This image was chosen for an initial study of filamentary structure because of the visual presence of filaments in a relatively uncomplicated area (Section \ref{sec:filaments}). This image has been smoothed, and the lines that mark the DR1 field edges have been minimized, to better highlight large scale structure.}
    \label{fig:fils}
\end{figure}

One outcome of the MDW Survey is its unique, high-resolution view of the Milky Way's warm, ionized diffuse gas and filamentary structure. Filaments have often been examined in neutral hydrogen, and combining these observations with their view in ionized \ha may provide insight into their alignment (or misalignment) with the Galactic magnetic field \citep{kim23, bally_jwst_hii,colin_smith_cloud}. To highlight the potential contribution of the MDW Survey in this scientific area, we present an initial study of filaments in the star-removed MDW image of the Lyra region, shown in Figure~\ref{fig:fils}.  

\begin{figure}
    \centering
    \includegraphics[width=1\linewidth]{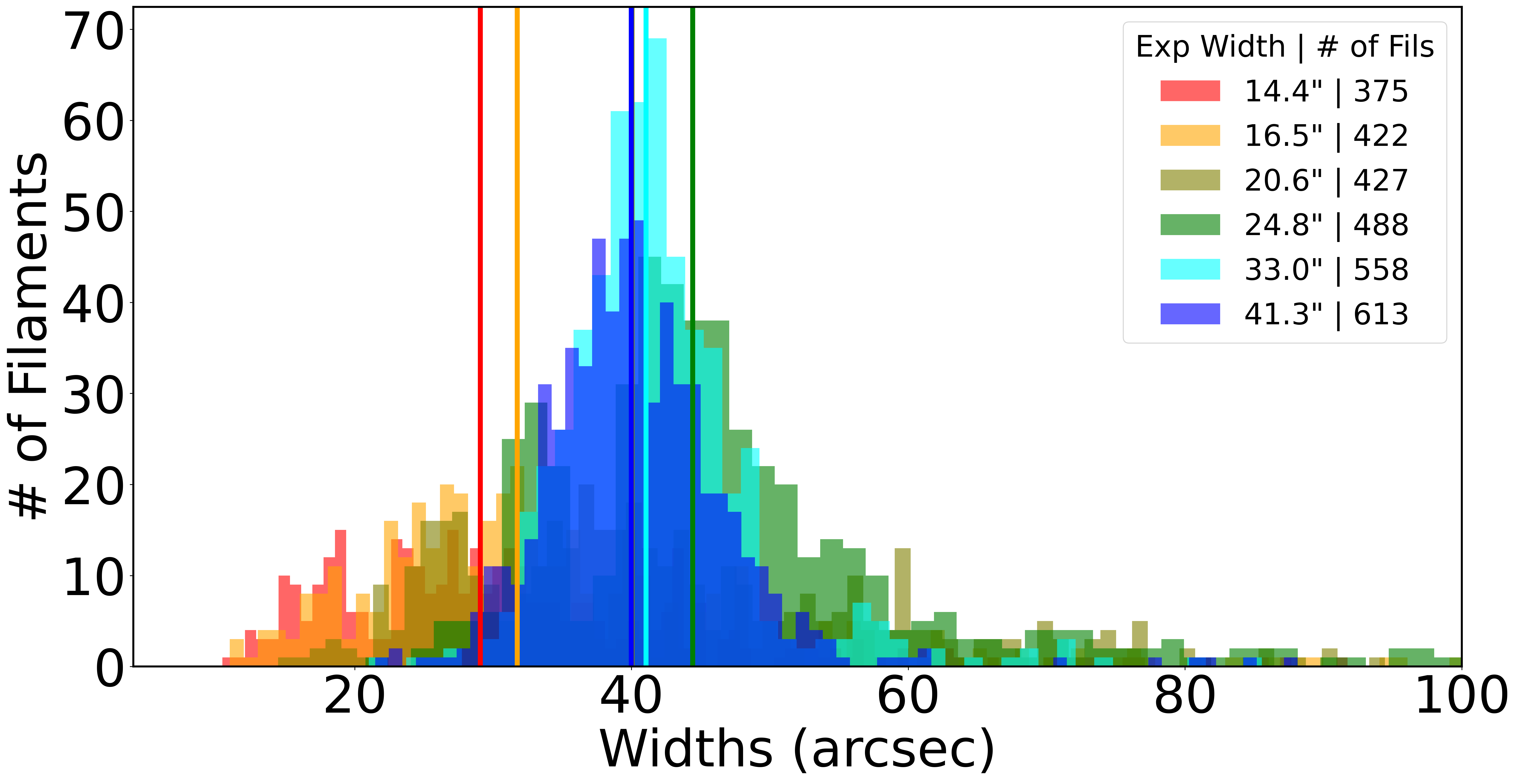}
    \caption{Distribution of filament widths, extracted from the full region immediately below and above the band of y-pixels containing Vega in Figure \ref{fig:fils} via \texttt{FilFinder}. We try a spread of expected widths as indicated by each color in the legend. Solid vertical lines in each color show the respective median width given the specified expected width. The number of filaments with successful fits and 1:2 aspect ratios are shown in the legend.}
    \label{fig:exp_widths_asp2}
\end{figure}

The Lyra region displays abundant linear filamentary features without excessive crowding from other ionized ISM structures, making it suitable for this initial study. To help detect filaments within Galactic emission, we first apply an unsharp mask (USM) to the image, with a 40\arcmin~Gaussian beam and a uniform Gaussian smoothing filter at $\sigma = 5$. 
We then use the \texttt{FilFinder} algorithm \citep{filfinder} to identify filaments, passing in a few key parameters. We set the scale width to approximately 6\arcsec, which is later used to set a smoothing size of 3\arcsec, and an adaptive threshold, relating to the expected filament width of 12\arcsec. The size threshold filters out filaments with total area smaller than the specified value, and is set to 1000 $\rm pix^2$. The globular (masking) threshold masks pixels with intensities smaller than the specified value, and is set to 15 Jy/pix. While we tried many variations on these parameters, the distribution of filament widths remained similar to those show in Figure~\ref{fig:exp_widths_asp2} (discussed in the next paragraph). We also use \texttt{FilFinder}'s border masking feature to avoid detecting filaments along, or cut off by, the image borders. The beamwidth is based on the MDW Survey's typical PSF FWHM, and is set to 6\arcsec.  We do not know the distance to the filamentary emission in the Lyra region, but we use 100 pc as an approximate lower limit based on the distance to the wall of the Local Bubble \citep{capitanio17, edenhofer2024, oneill2024}. 



Using the \texttt{FilFinder} parameters noted above, we were able to identify 
over 1000 structures in the regions immediately above and below the band of y-pixels containing Vega shown in Figure~\ref{fig:fils}. 
 We apply an aspect ratio cut of 1:2 to ensure the objects are somewhat elongated, which decreases the number of structures by a factor of $\sim2$.  We tried larger aspect ratios and found the median width of the filaments did not change significantly.
While we are able to extract multiple properties of these filaments, such as orientation, width, and intensity, we focus only on widths in this \ha~filament preview. The term 'width' is used to describe the FWHM of a Gaussian fit to the innermost part of a filament radial profile \citep{panopoulou2016}. The distribution of these extracted filament widths is shown in Figure~\ref{fig:exp_widths_asp2}. 
We exclude filaments that do not have a Gaussian fit for the specified expected width, flagged by \texttt{FilFinder}. Varying the expected width parameter can be used to choose between \texttt{FilFinder}'s ability to resolve the narrowest filaments versus its sensitivity to wider faint filaments. We find that changing this parameter shifts the median width and total number of filaments as depicted in Figure \ref{fig:exp_widths_asp2}. There is some evidence for the filament widths to peak around $\sim$30\arcsec$-$45\arcsec, for an expected width range of 14.4\arcsec\ to 41.3\arcsec. For smaller expected widths, we find that more filaments are flagged with a failed Gaussian fit, resulting in a smaller sample size. The number of filaments with successful width fits are noted in the legend of Figure \ref{fig:exp_widths_asp2}.

The measurement of filament width is resolution limited in most molecular, dusty or HI data \citep[][Putman et al.~2025]{panopoulou22}. The median width measured from the MDW \ha~data seems to be further resolved compared to these studies, as the typical filament FWHM is on the order of 40\arcsec and the MDW Survey's point sources typically have a 6\arcsec PSF. While we do not know the exact distance to these filaments, at the presumed distance of 100 pc, they would be 0.02 pc in width - this is much smaller than what is measured for filaments at other wavelengths.  
Future studies of \ha filaments in the MDW data will be a powerful way to characterize linear structures in the Galactic ISM.




\subsection{\ha variability and excess}
\label{sec:variability}

The MDW Survey is not optimally designed for the study of point-source \ha variability because observations are taken on an irregular cadence and time baseline (Figure \ref{fig:JD}; Section \ref{sec:cadence}). However, with up to twelve individual exposures of the same target field, we can still produce \ha variability catalogs, extract the objects with the most significant $\Delta$\ha, and with the help of external catalogs, broadly characterize them.

\begin{figure}
    \centering
    \includegraphics[width=1\linewidth]{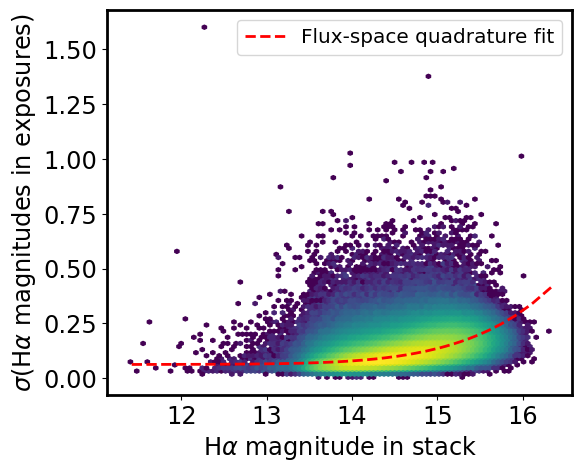}
    \caption{For sources in the DR1 stack catalog, their stack \ha magnitude compared to the standard deviation ($\sigma$) of their magnitudes across the individual exposures. The bins are colored by source density, revealing a quadrature relationship between stack \ha and $\sigma(\mathrm{H}\alpha)$. See Section \ref{sec:variability} for details.}
    \label{fig:ha_mag_std}
\end{figure}

Ideally, a static source tracked across multiple exposures will not change in its calibrated \ha magnitude. In reality, we expect slight fluctuations since the results of our point-source calibration can vary for exposures of different qualities (Section \ref{sec:reshoots}). We perform the same cuts on the DR1 stacked catalog as in Section \ref{sec:dp_catalogs_color_color}, and match the coordinates to the individual exposure catalogs (Section \ref{sec:dp_catalogs}) with a separation limit of 2\arcsec and checking that we were able to match the stack source in at least 80\% of the individual images\footnote{For this initial investigation of MDW \ha variability, we apply this condition to limit our sample to sources with good astrometry and signal-to-noise in both stacks and individual images.}. We calculate the standard deviation of each source's \ha magnitude (\sigha) over these multiple matched exposures, for 6 million stack sources . 

These results are plotted in Figure \ref{fig:ha_mag_std}, with bins shaded by the number of sources. We perform a quadrature fit in flux-space to better model the error, which is shown by the dashed red curve. The fainter a source is in the stack, the higher its \sigha\  across the individual frames, with the overall quadratic trend  attributed to calibration systematic uncertainties and the increased noise at fainter magnitudes. The trend line is relatively flat for magnitudes brighter than 14, at \sigha\ \textasciitilde0.1 AB magnitudes, and quickly increases for fainter stack sources, up to \textasciitilde0.3 AB magnitudes at m(\ha)$\sim$16. While the upper tail of the distribution may contain true variable sources, it is also possible that some sources are mismatched (Section \ref{sec:dp_catalogs}). 

\begin{figure*}
    \centering
    \includegraphics[width=1\linewidth]{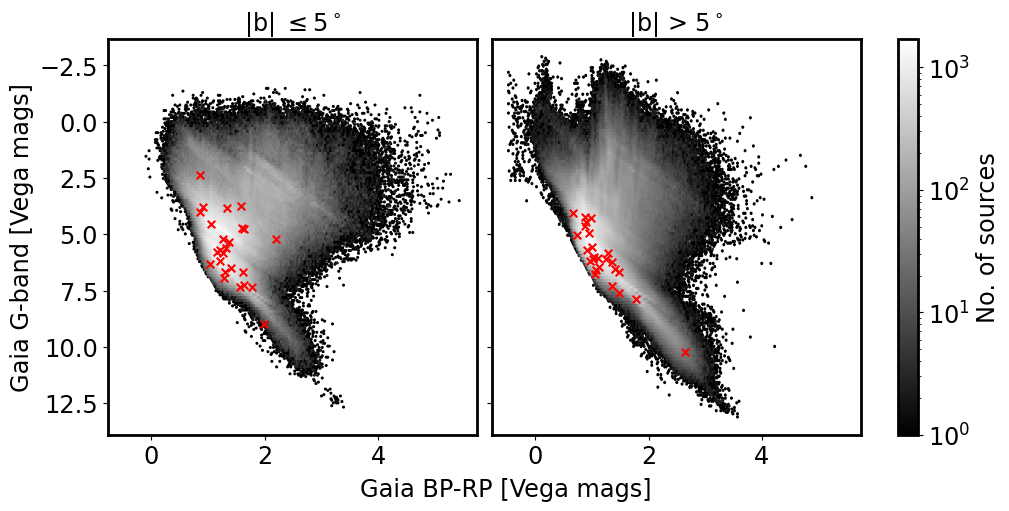}
    \caption{ The Gaia DR3 matches to the MDW Survey's DR1 stack objects presented in a HR diagram shaded by the number of sources. The 25 objects with the highest variation in their calibrated \ha magnitudes are overplotted in red crosses and further discussed in Section \ref{sec:variability}.}
    \label{fig:gaia_hr}
\end{figure*}

Figure \ref{fig:ha_mag_std} reveals numerous objects with \sigha $ \geq 0.8$ AB magnitude across the individual cutouts, which have $\geq 4\sigma$ variation in brightness relative to the median at the corresponding m(\ha). To understand their physical properties, we match to the Gaia Data Release 3 (DR3) catalog \cite{gaia_mission_2016b, gaia_dr3_summary_2023j}, with a radius of 1\arcsec. We apply the Gaia catalog cuts described in Appendix B of \cite{gaia_dr2_variable_hr}), resulting in approximately 10 million DR1-Gaia matches. Finally, we create a color-absolute magnitude diagram (CAMD; \cite{gaia_dr2_hr}), presented in Figure \ref{fig:gaia_hr} and section it by sources in the Galactic Plane ($|b| \leq 5^\circ$) and beyond it ($|b| > 5^\circ$). For both regions, we color the bins with a grayscale colorbar according to the number of sources, and overplot the 25 most variable sources using red crosses.  Nearly all these sources fall along the intermediate main sequence, possibly classifying them as accreting stars or binaries, although there may be a contribution from cooler, flaring red dwarves. Recent work has identified such candidates from IGAPS, Gaia, and S-PLUS \citep{Fratta2021Selection, Fratta2023Spec, SPHAS_Ha_EXCESS}. Further analysis is needed before we can definitively say what type of objects they are and what causes their \ha variability. 

It is also important to note that our algorithm for cosmic-ray detection (Section \ref{sec:cosmic_ray}) would flag singular \ha flares, from sources the survey cannot otherwise detect, as an unwanted artifact. These flares would not propagate into the final mean-combined image, and are also not accounted for in the mean-combined catalog. While we expect this to be a small number of events, a thorough and complete study of variable objects in the MDW Survey requires us to account for flares, and is on the roadmap for the DR2 release.

\begin{figure}
    \centering
    \includegraphics[width=1\linewidth]{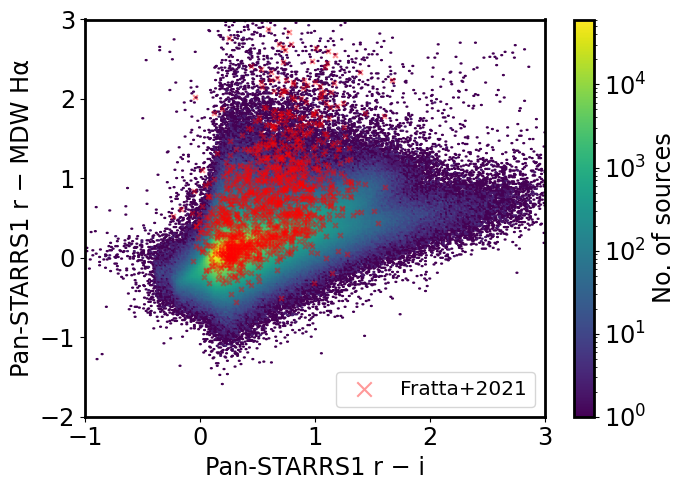}
    \caption{For DR1 stack matches to Pan-STARRS1, the $r-\mathrm{H}\alpha$ vs. $r-i$ color-color plane, with the matches to the catalog produced by \cite{Fratta2021Selection} overplotted in red crosses. We limit our matches with \cite{Fratta2021Selection} to 5$\sigma~ $\ha outliers, revealing many probably \ha-excess objects and a handful of sources showing \ha-absorption.}
    \label{fig:fratta}
\end{figure}

Finally, we search for possible \ha excess objects in DR1 by comparing the DR1 stack catalog to a catalog of potential \ha-excess objects produced by \cite{Fratta2021Selection}. We match our Pan-STARRS1 coordinates to their given Gaia coordinates with a separation limit of 0.12\arcsec. We filter their catalog to sources with 5$\sigma$-certainties, using the \texttt{flagPOS} and \texttt{flagCAMD} properties - these indicate the certainty of IGAPS \ha-excess derived from the Gaia position-based criteria and CAMD-based criteria described in \cite{Fratta2021Selection}. This results in a sample of 1500 objects. We overlay the positions of these matches on our $r-\mathrm{H}\alpha$ vs $r-i$ color-color plane, applying the same conditions given in Section \ref{sec:dp_catalogs_color_color}. Figure \ref{fig:fratta} shows these matches overplotted in red crosses. We find many of these matches sitting within the stellar locus, where most sources reside, and along the unreddened main sequence. Notably, we see numerous points falling far above the unreddened main sequence, evenly scattered along the $ 1 \leq r-\mathrm{H}\alpha \leq 3$ range. Section \ref{sec:dp_catalogs} discusses the MDW-Pan-STARRS1 mismatches contaminating this range of colors, but we see this spread can also be attributed to the presence of real \ha-excess sources. Since the MDW Survey's \ha filter is narrower than IGAPS, we expect to be more sensitive to \ha excess and absorption, skewing our $r-\mathrm{H}\alpha$ values higher.

\section{Summary}
\label{sec:conclusion}

The MDW \ha Sky Survey is a single-band, full-sky imaging survey using a narrow \ha filter centered at 656.3 nm with a bandwidth of 3 nm. Building on our first Data Release of the Orion region \citep{MDW_DR0_paper}, we now present Data Release 1 (DR1) of the MDW Survey, spanning the full Northern Sky (Dec $\geq 0$, or 50\% of the entire sky). 

Since DR0, we have improved various elements of our data reduction pipeline, fully migrating to using open-source software, expanded our coverage of the sky ten-fold (5\% in DR0 vs. 50\% in DR1), and have extended our pipeline to include our multi-epoch individual exposures. We fetch synthetic main sequence spectra to compare their theoretical magnitudes through the MDW Survey's \ha filter to the broader IGAPS \ha filter, finding that the narrower MDW filter should be more sensitive to \ha emission and absorption.


We release images for each field, of the mean-combined stack, their star-removed counterpart, and the individual epoch exposures. We also include point-source catalogs for both the stacks and exposures, matched to Pan-STARRS1 and IGAPS to provide valuable color information. We plan on improving our astrometry in Data Release 2 (DR2) to increase match accuracy - in the meantime, we give a set of conditions users can apply to filter the catalogs to sources with good Pan-STARRS1 matches and good signal-to-noise. All DR1 data products are outlined in Table \ref{tab:data_products}. 

The MDW Survey provides a new avenue to examine diffuse \ha gas and point-source \ha activity in the northern sky, and we perform various initial science work using the DR1 dataset. We use star-removed images to examine ionized hydrogen filaments in the Lyra region, extracting their widths using \texttt{FilFinder}. We begin investigating \ha variability in point sources, and use the Gaia CAMD to broadly characterize the most variable sources. We also match the DR1 stack catalog to the potential \ha-excess sources given in \cite{Fratta2021Selection}, and find that many can present with MDW \ha magnitudes brighter than Pan-STARRS1 r by up to \textasciitilde3 AB magnitudes.

The MDW Survey expects to complete observations of the full sky by early 2026, and we will proceed to release our DR2 dataset later that year.
In addition to the point-source flux calibration done in DR0 and DR1, we plan to implement a diffuse emission calibration by comparing to the \ha composite map presented in \cite{finkbeiner_2003}. We are also preparing our data to be archived in MAST. The MDW Survey will be a legacy all-sky \ha data source for all to utilize.

\section*{Acknowledgments}

Funding for the MDW Survey Project has been provided by the Michele and David Mittelman Family Foundation. David R. Mittelman, Dennis di Cicco, and Sean Walker are founding members of the survey and made possible the acquisition and reduction of the data. The Columbia University Astronomy Department is responsible for the final data reduction, calibration, and dissemination of the survey data.

s

We are deeply thankful to Katherine Blundell for her operational support of the MDW Survey. We are also very grateful to David Sliski for his early management and direction of the MDW Survey, and to Arne Henden and Gary Walker, for their support in the MDW Survey's observing operations.

We extend our appreciation to the following undergraduate and post-baccalaureate researchers, who have all worked to push forward the MDW Survey datasets and its early scientific results: Julia Homa, Colin Holm-Hansen, Lylon Sanchez Valido, and Joshua Thorne. 


The Pan-STARRS1 Surveys (PS1) and the PS1 public science archive have been made possible through contributions by the Institute for Astronomy, the University of Hawaii, the Pan-STARRS Project Office, the Max-Planck Society and its participating institutes, the Max Planck Institute for Astronomy, Heidelberg and the Max Planck Institute for Extraterrestrial Physics, Garching, The Johns Hopkins University, Durham University, the University of Edinburgh, the Queen's University Belfast, the Harvard-Smithsonian Center for Astrophysics, the Las Cumbres Observatory Global Telescope Network Incorporated, the National Central University of Taiwan, the Space Telescope Science Institute, the National Aeronautics and Space Administration under Grant No. NNX08AR22G issued through the Planetary Science Division of the NASA Science Mission Directorate, the National Science Foundation Grant No. AST-1238877, the University of Maryland, Eotvos Lorand University (ELTE), the Los Alamos National Laboratory, and the Gordon and Betty Moore Foundation.

DR1 makes use of data obtained as part of the IGAPS merger of the IPHAS and UVEX surveys (\url{www.igapsimages.org}) carried out at the Isaac Newton Telescope (INT). The INT is operated on the island of La Palma by the Isaac Newton Group in the Spanish Observatorio del Roque de los Muchachos of the Instituto de Astrofisica de Canarias. All IGAPS data were processed by the Cambridge Astronomical Survey Unit, at the Institute of Astronomy in Cambridge.

This work has made use of data from the European Space Agency (ESA) mission
{\it Gaia} (\url{https://www.cosmos.esa.int/gaia}), processed by the {\it Gaia}
Data Processing and Analysis Consortium (DPAC,
\url{https://www.cosmos.esa.int/web/gaia/dpac/consortium}). Funding for the DPAC
has been provided by national institutions, in particular the institutions
participating in the {\it Gaia} Multilateral Agreement.

This research made use of ccdproc, an Astropy package for image reduction. 

This research made use of Photutils, an Astropy package for
detection and photometry of astronomical sources \citep{photutils}.

This work made use of Astropy:\footnote{\url{http://www.astropy.org}} a community-developed core Python package and an ecosystem of tools and resources for astronomy \citep{astropy}.

This work made use of Montage\footnote{\url{http://montage.ipac.caltech.edu/}}.It is funded by the National Science Foundation under Grant Number ACI-1440620, and was previously funded by the National Aeronautics and Space Administration's Earth Science Technology Office, Computation Technologies Project, under Cooperative Agreement Number NCC5-626 between NASA and the California Institute of Technology.

We acknowledge the use of large-language models (LLMs) to debug and enhance the code generating the DR1 data products, and to assist in plot creation.


\software{Astropy \citep{astropy:2013, astropy:2018, astropy:2022}, photutils \citep{photutils}, Matplotlib \citep{matplotlib}, NumPy \citep{numpy}, SciPy \citep{scipy}, Pandas \citep{pandas}, astrometry.net \citep{astrometry_net}, MontagePy \citep{montagepy}, ccdproc \citep{ccdproc}, CloudClean \citep{saydjari}, MESA MIST \citep{MIST_1, MIST_2}, synphot \citep{synphot}, FilFinder \citep{filfinder}}

\newpage
\appendix

\section{Supplementary data on catalogs and synthetic photometry}
\label{appendix:ps1_flags}
\label{appendix:cat_phot_supp}

\subsection{Pan-STARRS1 matching}

Before matching our DR1 catalogs to the Pan-STARRS1 survey, we access the Pan-STARRS DR1 catalog using the MAST CasJobs\footnote{\url{https://mastweb.stsci.edu/ps1casjobs/home.aspx}} interface, filtering for objects with at least 2 detections in the r filter. The exact SQL query is given in Listing 1.

\lstset{
  captionpos=b,
  language=SQL,
}
\begin{lstlisting}[ language=SQL, caption=The SQL query we pass into MAST CasJobs to fetch Pan-STARRS1 DR1 objects for MDW Survey catalog-matching. ]
select o.objID, o.nDetections, o.raMean, o.decMean, o.raMeanErr, o.decMeanErr,
       m.gMeanPSFMag, m.gMeanPSFMagErr, m.rMeanPSFMag, m.rMeanPSFMagErr,
       m.iMeanPSFMag, m.iMeanPSFMagErr, o.qualityFlag, o.objInfoFlag,
       m.rFlags, m.gFlags, m.iFlags
from ObjectThin o
inner join MeanObject m on o.objID = m.objID
where
    o.decMean >= -5 and
    o.nDetections > 1 and
    o.nr > 2 and
    m.rMeanPSFMag <= 18 
\end{lstlisting}

When calibrating our point-sources (Section \ref{sec:psm_flux_calibration}) and constructing color-color diagrams (e.g. in Figure \ref{fig:color_color}), we only use Pan-STARRS1 matches that pass the \texttt{passes\_ps\_quality\_check} condition. The Pan-STARRS1 catalog flags that comprise our \texttt{passes\_ps\_quality\_check} flag is outlined in Table \ref{table:ps_flags}.

\begin{table}[H]
\centering
\caption{Pan-STARRS1 DR1 bitmask flags used to refine magnitude calibration cuts in Section \ref{sec:psm_flux_calibration}, and used when displaying color-color diagrams (Section \ref{sec:dp_catalogs_color_color}). Pan-STARRS1 sources are excluded if any of the flags below fail the indicated condition.}
\begin{tabular}{lll}
\toprule
\textbf{Pan-STARRS1 Table} & \textbf{Flag(s)} & \textbf{Condition} \\
\texttt{ObjectThin}   & \texttt{qualityFlag}   & \begin{tabular}[t]{@{}l@{}}
\texttt{!QF\_OBJ\_EXT} \&  \\
\texttt{QF\_OBJ\_GOOD\_STACK} \& \\
\texttt{QF\_OBJ\_GOOD} \& \\
\texttt{!QF\_OBJ\_SUSPECT\_STACK} \& \\
\texttt{!QF\_OBJ\_BAD\_STACK} \\

\end{tabular} \\ \\ 
\texttt{MeanObject}   & \texttt{rFlags}, \texttt{iFlags}, \texttt{gFlags} & 
\begin{tabular}[t]{@{}l@{}}
\texttt{SECF\_RANK\_0} \&  \\
\texttt{!SECF\_STAR\_FEW} \& \\
\texttt{!SECF\_STAR\_POOR} \\
\end{tabular} \\
\bottomrule
\end{tabular} 
\label{table:ps_flags}
\end{table}
\begin{figure}
\end{figure}

\subsection{Synthetic photometry}
\begin{figure}
    \centering
    \includegraphics[width=0.75\linewidth]{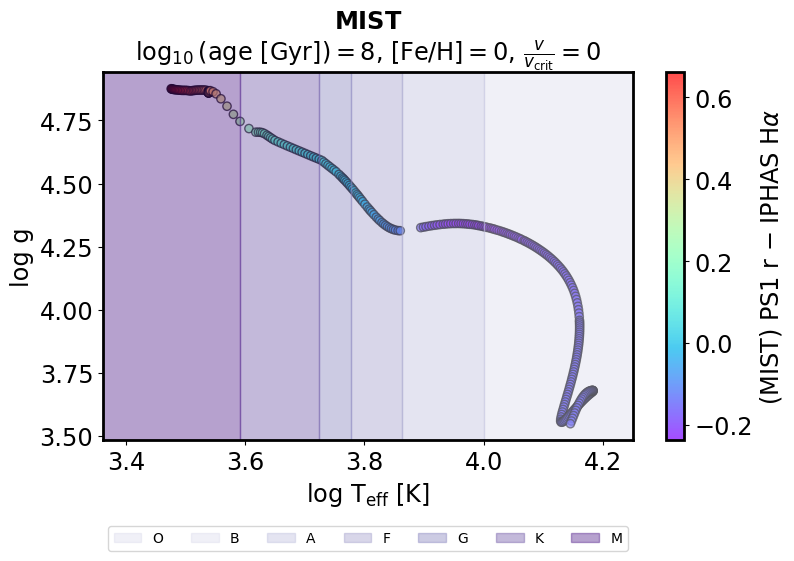}
    \caption{The MIST data points, which we compare to at various points in this manuscript to the MDW-PS1 and PS1-IGAPS color-color plane (Sections \ref{sec:psm_flux_calibration}, \ref{sec:dp_catalogs_color_color}). We believe these data points correspond to the unreddened main sequence, and refer to it as such.}
    \label{fig:mist}
\end{figure}

\begin{figure}
    \centering
    \includegraphics[width=0.9\linewidth]{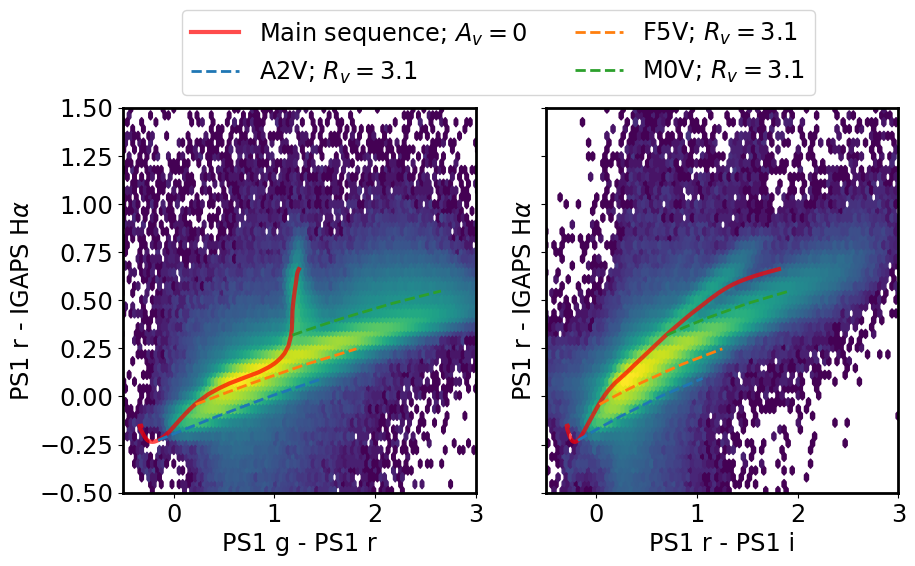}
    \caption{Similar to Figure \ref{fig:color_color}, but instead matching the full Pan-STARRS1 DR1 catalog to IGAPS, we can create reference color-color planes. From these, we can compare how the color-color plots matching Pan-STARRS1 to the MDW Survey differ from the "standard", and how much deviation to expect in the synthetic MIST tracks. }
    \label{fig:ps_igaps}
\end{figure}

At various points in this paper (Section \ref{sec:psm_flux_calibration}; Section \ref{sec:dp_catalogs_color_color}), we compare our DR1 colors to the MIST synthetic tracks provided for the Pan-STARRS1 and IGAPS surveys. These MIST tables are generated for various stellar physical properties, so we perform certain cuts to narrow down the synthetic Pan-STARRS1-IGAPS data points to the unreddened main sequence. Specifically, we filter the MIST tables to solar-metallicity stars with no surface velocity ($\frac{v}{v_{\mathrm{crit}}}=0$) and with $\log_{10}{\mathrm{age [Gyr]}}= 8$. These parameters select the unreddened main sequence, supported by the appropriate range in surface gravity and effective temperatures of the MIST data points shown in Figure \ref{fig:mist}. Figure \ref{fig:ps_igaps} demonstrates the general agreement of these tracks between matched Pan-STARRS1 and IGAPS sources, reassuring its use as a reference for the MDW point-source flux calibration and when overlaying the tracks in our color-color diagram.

\section{FWHM Refitting}
\label{appendix:fwhm_refitting}

We find that the \texttt{photutils} PSF photometry API may exaggerate the fitted FWHM for some sources. This effect is more pronounced in crowded fields, and could be due to low signal-to-noise or proximity to other bright sources skewing the FWHM fitting. We choose to refit the FWHM to an exponential function, using the instrumental magnitudes as the primary input. This is a per-field fitting algorithm.


\begin{figure}[t] 
    \centering
    \includegraphics[width=0.6\linewidth]{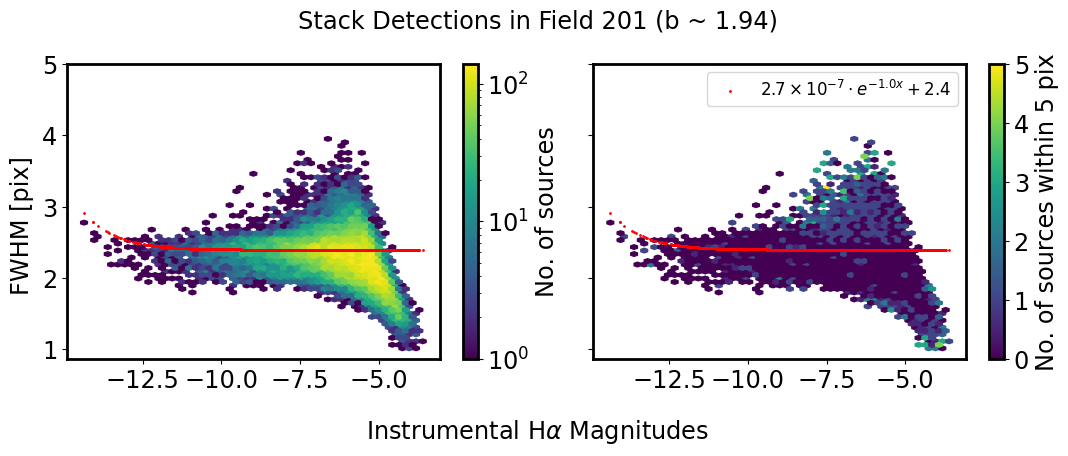}
    \includegraphics[width=0.6\linewidth]{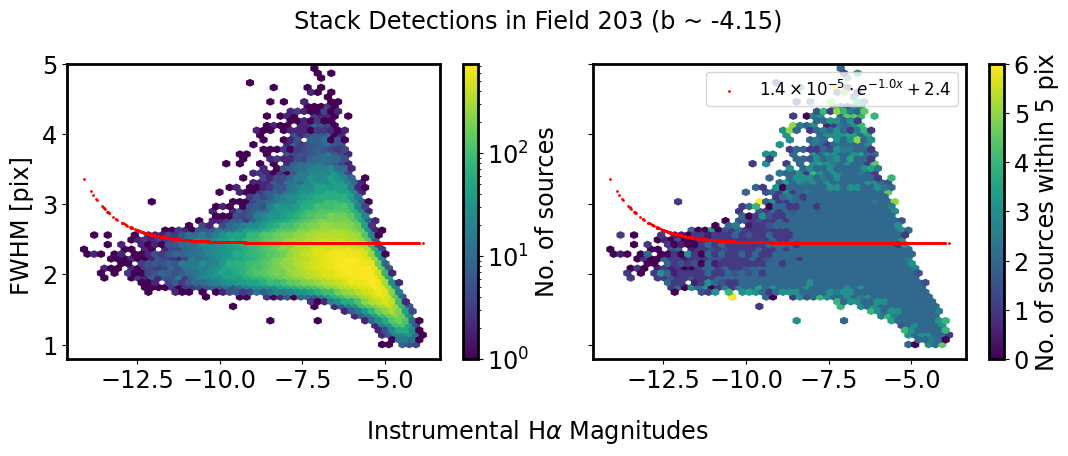}
    \includegraphics[width=0.6\linewidth]{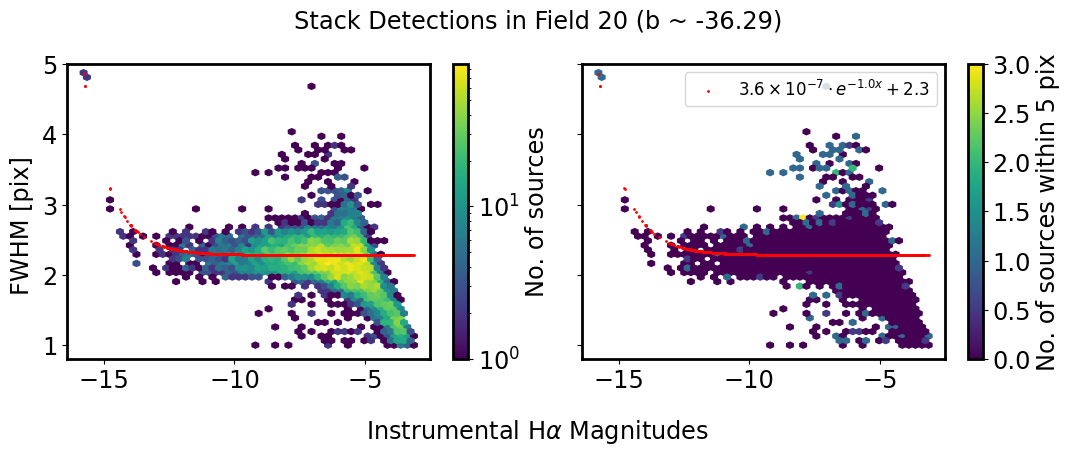}
    \includegraphics[width=0.6\linewidth]{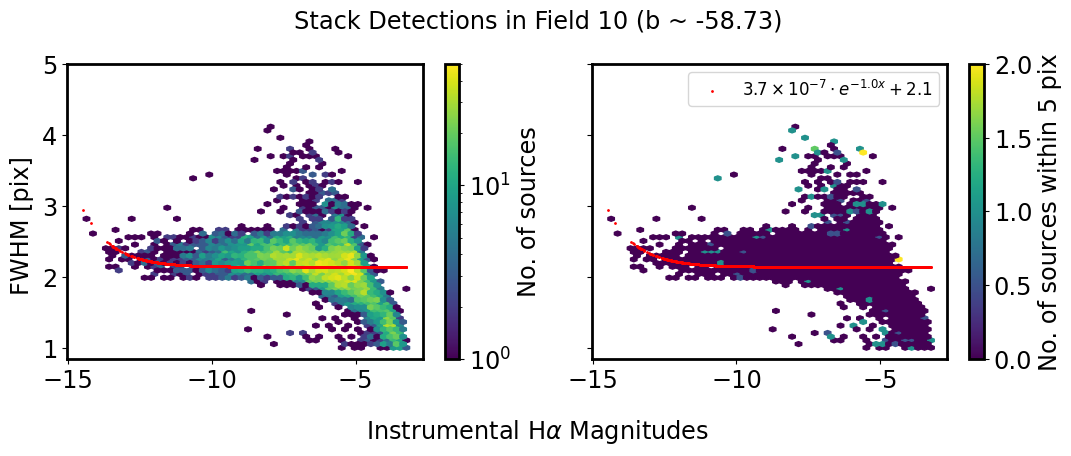}
    \caption{Per-field examples of refitting source FWHMs to an exponential, aiding in the source masking process in Section~\ref{sec:star_removal}.}
    \label{fig:fwhm_refitting}
\end{figure}

\clearpage

\section{Catalog Columns}
\label{appendix:c}
\label{appendix:catalog_columns}

For the DR1 stack and individual exposure source catalogs, we provide the name of each column provided, its data type, and its description in Table \ref{tab:catalog_columns}.

    \begin{longtable}{|l|l|p{10cm}|}
        \caption{List of columns in the MDW Survey's DR1 point source catalogs, matched to the Pan-STARRS1 DR1 and IGAPS survey catalogs. } 
        \label{tab:catalog_columns}
        \\
        
        \hline
        \textbf{Column} & \textbf{Type} & \textbf{Description} \\ \hhline{|=|=|=|}
        
        {MDW\_RAJ2000} & float64 & MDW J2000 Right Ascension (degrees) \\ \hline
        
        {MDW\_DEJ2000} & float64 & MDW J2000 Declination (degrees) \\ \hline

        {MDW\_l} & float64 & MDW Galactic longitude (degrees) \\ \hline
        
        {MDW\_b} & float64 & MDW Galactic latitude (degrees) \\ \hline
        
        {MDW\_Ha\_mag} & float64 & MDW calibrated \ha AB magnitude \\ \hline
        
        {MDW\_Ha\_instr\_mag} & float64 & MDW instrument \ha magnitude, calculated from MDW\_Ha\_instr\_flux \\ \hline
        
        {MDW\_Ha\_instr\_flux} & float64 & MDW instrument flux via PSF photometry \\ \hline

        {fitted\_fwhm} & float64 & Source FWHM, fitted to an exponential according to the instrument magnitude (Section \ref{sec:star_removal}; Appendix \ref{appendix:fwhm_refitting}). \\ \hline
        
        {field} & int32 &  Field number in which the source is found in. \textsuperscript{a}\\ \hline

        {file\_name} & |S20 &  Name of the FITS file this source is found in. \textsuperscript{a}\\ \hline
        
        {ps\_RAJ2000} & float64 & Pan-STARRS1 J2000 Right Ascension (degrees) \\ \hline
        
        {ps\_DEJ2000} & float64 & Pan-STARRS1 J2000 Declination (degrees) \\ \hline
        
        {ps\_objID} & int64 & Pan-STARRS1 object ID \\ \hline
        
        {ps\_g\_mag} & float64 & Pan-STARRS1 g AB magnitude \\ \hline
        
        {ps\_r\_mag} & float64 & Pan-STARRS1 r AB magnitude \\ \hline
        
        {ps\_i\_mag} & float64 & Pan-STARRS1 i AB magnitude \\ \hline
        
        

        {n\_ps\_nearby} & int32 &  Number of Pan-STARRS1 objects within a 6.0" radius of the MDW source. \\
        \hline
        
        {ig\_RAJ2000} & float64 &  IGAPS J2000 Right Ascension (degrees) \textsuperscript{c} \\ \hline
        
        {ig\_DEJ2000} & float64 & IGAPS J2000 Declination (degrees) \textsuperscript{c}  \\ \hline
        
        {ig\_name} & |S20 & IGAPS name \textsuperscript{c} \\ \hline
        
        {ig\_r\_mag} & float64 & IGAPS r AB magnitude \textsuperscript{c}  \\ \hline
        
        {ig\_i\_mag} & float64 & IGAPS i AB magnitude \textsuperscript{c}  \\ \hline
        
        {ig\_g\_mag} & float64 & IGAPS g AB magnitude \textsuperscript{c}  \\ \hline
        
        {ig\_Ha\_mag} & float64 & IGAPS \ha AB magnitude. \textsuperscript{c}  \\ & & \\ & & If present, this should be similar to MDW\_Ha\_mag (see Figure \ref{fig:stack_colors}). \\ \hline
                
        {MDW\_x} & float64 & The x coordinate where the MDW source is found in the field image \\ \hline
        
        {MDW\_y} & float64 & The y coordinate where the MDW source is found in the field image \\ \hline
        
        {mdw\_ps\_sep2d} & float64 &  The 2D separation distance between the MDW source and the matched Pan-STARRS1 source  \\ \hline

        {ps\_igaps\_sep2d} & float64 &  The 2D separation distance between the matched Pan-STARRS1 source and IGAPS \textsuperscript{c} \\ \hline

        {avg\_date} & |S20 &  This column only exists for individual catalogs. The average date/time of the exposure this individual source is found in. \textsuperscript{d}\\ \hline

    \end{longtable}

    \begin{tablenotes}
            \small 
            \item{$^\text{a}$ \footnotesize In practice, a source can be found in two fields if it is located in the field overlap region. }

            \item{$^\text{sb}$ \footnotesize A source's ultimate match is to the closest Pan-STARRS1 object within a 1.5" radius. However, we note the number of Pan-STARRS1 objects within a 6" radius to help account for the possibility of mismatches (see Section \ref{sec:dp_catalogs_color_color}). }

            \item{$^\text{c}$ \footnotesize If there's no matched IGAPS source, this value is -999.
            
            \item{$^\text{d}$ \footnotesize Individual exposures are 20 minutes long in the MDW Survey.}}
        \end{tablenotes}
\clearpage

\clearpage

\bibliographystyle{aasjournal}
\bibliography{dr1}

\end{document}